\documentclass[aoas,MSNbibl,nameyear,seceqn,dvips]{arximspdf}
\usepackage{multirow}
\usepackage{graphicx}

\doi{10.1214/12-AOAS567} 
\volume{6}
\issue{4}
\pubyear{2012}
\firstpage{1861}
\lastpage{1882}



\begin{document}
\begin{frontmatter}

\title{A semiparametric regression model for paired longitudinal outcomes
with application in childhood blood pressure development\thanksref{T1}}
\thankstext{T1}{Supported by Grant R01 HL095086 from the National Heart,
Lung, and Blood Institute, United States Department of Health and Human
Services.}
\runtitle{Semiparametric model for paired longitudinal outcomes}

\begin{aug}
\author[A]{\fnms{Hai} \snm{Liu}\corref{}\ead[label=e1]{liuhai@iupui.edu}} 
\and
\author[A]{\fnms{Wanzhu} \snm{Tu}\ead[label=e2]{wtu1@iupui.edu}}

\runauthor{H. Liu and W. Tu}
\affiliation{Indiana University School of Medicine}
\address[A]{Department of Biostatistics\\
Indiana University School of Medicine\\
Indianapolis, Indiana 46202\\
USA\\
\printead{e1}\\
\phantom{E-mail:\ }\printead*{e2}} 
\end{aug}

\received{\smonth{7} \syear{2011}}
\revised{\smonth{5} \syear{2012}}

%
\begin{abstract} 
This research examines the simultaneous influences of height and weight
on longitudinally measured systolic and diastolic blood pressure in
children. Previous studies have shown that both height and weight are
positively associated with blood pressure. In children, however, the
concurrent increases of height and weight have made it all but
impossible to discern the effect of height from that of weight. To
better understand these influences, we propose to examine the joint
effect of height and weight on blood pressure. Bivariate thin plate
spline surfaces are used to accommodate the potentially nonlinear
effects as well as the interaction between height and weight. Moreover,
we consider a joint model for paired blood pressure measures, that is,
systolic and diastolic blood pressure, to account for the underlying
correlation between the two measures within the same individual. The
bivariate spline surfaces are allowed to vary across different groups
of interest. We have developed related model fitting and inference
procedures. The proposed method is used to analyze data from a real
clinical investigation.
\end{abstract}

%
\begin{keyword}
\kwd{Bootstrap}
\kwd{factor-by-surface interaction}
\kwd{mixed effects model}
\kwd{paired outcomes}
\kwd{penalized estimation}
\kwd{thin plate spline}.
\end{keyword}

\end{frontmatter}

\section{Introduction}
Excess weight gain has long been recognized as a risk factor for
metabolic and cardiovascular disorders, including hypertension.
Population studies have shown that weight strongly predicts blood
pressure [\citet{Huang1998}, \citet{Masuo2000}], although the relationship
between the two may not be linear [\citet{Hall2010}]. Data from children
and young adults are equally persuasive on the weight--blood pressure
association [\citet{Stray2009}, \citet{Levin2010}]. In fact, the observations
are so consistent that some even question whether obesity and
hypertension are two epidemics or one [\citet{Davy2004}]. More recently,
an increasing number of studies have recognized a similarly salient
relationship between height and blood pressure [\citet{Shankar2005}, \citet{Fujita2010}].
Although this latter relationship appears to have a
seemingly plausible physiological interpretation (taller individuals
need greater pressure to maintain oxygenated blood flow to the head and
upper extremities), the concurrent increases of height and weight have
nonetheless made it analytically difficult to discern the effect of
weight from that of height. The matter is further complicated by the
observed significantly positive effect of body mass index (BMI, defined
as Weight$/$Height$^{2}$) on blood pressure, as shown in numerous studies
[\citet{Lauer1989}, \citet{Baker2007}, \citet{Falkner2010}]. In fact, overweight and
obesity, clinically defined by BMI cutoff points, have been recognized
as major risk factors for hypertension, and, indeed, hypertension
prevalence is much higher in overweight and obese children [\citet
{Falkner2006}, \citet{Steinberger2009}]. Therefore, the scientific community
has a great interest to elucidate the independent influences of height
and weight on blood pressure, as they may implicate different
pathophysiology for this etiologically less understood disease. For
example, an obesity mediated blood pressure elevation would implicate a
more activated sympathetic nervous system (perhaps stimulated by
adipose-derived hormones such as leptin), and increased sodium
reabsorption by the kidney [\citet{Hall2010}], whereas a strong height
influence could give more credence to the notion that the disease has
its origin in growth [\citet{Lever1992}].

In this research, we assess the simultaneous influences of height and
weight on blood pressure using prospectively collected data from a
cohort of healthy children. Such an exercise, however, faces a number
of methodological challenges: (1)~Outcomes are \textit{repeatedly measured
paired} observations. Blood pressure consists of two readings, a
systolic measurement taken during the contraction phase of the cardiac
cycle and a diastolic measurement taken during the recoil phase of the
cardiac cycle. Together, they represent the pressure exerted by the
circulating blood on the walls of blood vessels during two different
phases of the same cardiac cycle. For longitudinally collected blood
pressure measurements, separate modeling of the systolic and diastolic
outcomes may not be appropriate, as the two are biologically correlated
and mutually influential [\citet{Guo2004}]. (2)~Height and weight
effects on blood pressure may be nonlinear. Previous research has
recognized a nonlinear pattern of the adiposity effects on blood
pressure [\citet{Hall2003}]. More recent data indicate a nonlinear
height effect on blood pressure as well [\citet{Tu2009}]. Preliminary
data from this investigation also indicate nonlinear height and weight
effects on blood pressure; see Figure~\ref{figmarginal} in Section~\ref
{seccase}. Furthermore, interaction of height and weight may exist. To
accommodate, \textit{nonlinear bivariate effect surfaces} will have to be
incorporated into the model for the purpose of depicting the joint
height--weight effects on blood pressure. (3)~Inferences are needed for
comparing the joint height--weight effects across different gender and
ethnicity groups. Such comparisons are of great interest, as recent
reports indicate significantly higher risks of obesity and hypertension
in black children than in white children [\citet{Anderson2009}, \citet{Brady2010}].
Differentially expressed bivariate effect surfaces,
therefore, may well point to different pathophysiology of the disease
among people of different ethnic backgrounds.

To address these challenges, we propose a joint semiparametric mixed
effects model that includes two individual components, one for systolic
and the other for diastolic blood pressure measures. Bivariate smooth
functions for the joint height--weight effects are embedded in these
components to account for possible nonlinear effects as well as
interactions between the two independent variables. The two components
are then connected by shared random subject effects in a unified
regression framework.

Semiparametric regression as a practical data analytical tool has
experienced tremendous growth in the past ten years, especially since
the publication of the book \textit{Semiparametric Regression} by \citet
{Ruppert2003}. Methodological extensions, stimulated by exciting
applications, and new computational approaches, have now covered most
commonly encountered data situations. \citet{He2005} considered robust
generalized estimating equations (GEE) for analyzing longitudinal data
in generalized partial linear models. \citet{Lin2006} proposed profile
kernel and backfitting estimation methods for a class of semiparametric
problems. Models with bivariate smoothing have been developed for
geospatial [\citet{Sain2006}, \citet{Guillas2010}] and medical imaging
applications [\citet{Brezger2007}]. Penalized splines have been used to
analyze longitudinally measured event counts [\citet{Dean2007}]. \citet
{Crain2008} have proposed penalized bivariate splines for binary
response and developed related Bayesian inference procedures. More
recently, \citet{Ghosh2009} have developed a joint semiparametric
structure for zero-inflated counts that consists of a logistic model
for the proportion of zeros and a log-linear model for Poisson counts;
both models contain univariate nonparametric components. \citet
{Ghosh2010} studied a semiparametric model for multivariate
longitudinal data in a Bayesian framework. Although there is a rich
literature on semiparametric analysis of longitudinal (or clustered)
data, not much has been developed for analyzing bivariate joint effects
of two continuous independent variables (height and weight in this
application) on a pair of closely related outcome variables (e.g.,
systolic and diastolic blood pressure).
The current research extends the previous work by introducing
group-specific bivariate smooth components into the joint modeling of
paired outcomes. Related model fitting and inference procedures are
also developed.

The outline of this paper is as follows. We introduce the semiparametric
mixed model for paired outcomes and its estimation in Section~\ref{secmethods}.
Hypothesis testing for group differences in the bivariate effect
surfaces is discussed in
Section~\ref{sectesting}, followed by a Monte Carlo study in Section
\ref{secsim}.
As the motivation and illustration of the proposed methods,
a real data application of childhood blood pressure study is presented
in Section~\ref{seccase}. We conclude the paper with a few
methodological and scientific remarks in Section~\ref{secdiscussion}.
Additional details on model-fitting algorithms and model diagnostics
are provided in the supplementary materials [\citet{LiuTu2012S}].

\section{Methods}
\label{secmethods}

\subsection{A semiparametric mixed model for paired outcomes}

We introduce our model in a more generic setting. Let $\mathbf
{Y}_{ij}= (Y^{(1)}_{ij},Y^{(2)}_{ij} )^{T}$ be a pair of outcomes
from the $i$th subject measured at the $j$th visit,
where $j=1,\ldots,n_i$ and $i=1,\ldots,m$. Assuming that we have
$g=1,\dots,G$ groups of interest,
let $z_{ig}$ be a binary group indicator for the $i$th subject:
$z_{ig}=1$ if the $i$th subject
belongs to the $g$th group, $z_{ig}=0$ otherwise. We propose the
following semiparametric mixed effects model
for the paired outcomes:
%
\begin{equation}
\label{eqjoint} \cases{ %
\displaystyle Y^{(1)}_{ij}
= U^{(1)}_{i} + \mathbf{t}_{ij}^{T}
\bolds{\psi}_1 + \sum_{g=1}^{G}
f^{(1)}_{g}(w_{ij}, h_{ij})z_{ig}+
\epsilon^{(1)}_{ij}, \vspace*{2pt}
\cr
\displaystyle
Y^{(2)}_{ij} = U^{(2)}_{i} +
\mathbf{t}_{ij}^{T}\bolds{\psi}_2 + \sum
_{g=1}^{G} f^{(2)}_{g}(w_{ij},
h_{ij})z_{ig}+\epsilon^{(2)}_{ij}, }
\end{equation}
where $\widetilde{\mathbf{U}}_i= (U^{(1)}_{i}, U^{(2)}_{i} )^{T}$
is the random subject effect vector;
$\mathbf{t}_{ij}$ denotes the time-varying covariate vector of the
$i$th subject at visit $j$
whose effects are assumed to be parametric with corresponding parameter
vectors $\bolds{\psi}_1$
and $\bolds{\psi}_2$; the joint influences of two continuous
independent variables
$w_{ij}$ and $h_{ij}$ in each group are modeled by the group-specific
bivariate smooth functions
$f^{(1)}_{g}$ and $f^{(2)}_{g}$ for the paired outcomes respectively;\vspace*{-1pt}
$\epsilon^{(1)}_{ij}$ and $\epsilon^{(2)}_{ij}$ are random errors.
Herein, we let $\mu^{(1)}_{ij}=\mathbf{t}_{ij}^{T}\bolds{\psi}_1 + \sum_{g=1}^{G} f^{(1)}_{g}(w_{ij}, h_{ij})z_{ig}$ and
$\mu^{(2)}_{ij}=\mathbf{t}_{ij}^{T}\bolds{\psi}_2 + \sum_{g=1}^{G}
f^{(2)}_{g}(w_{ij}, h_{ij})z_{ig}$ to denote the mean responses.

The regression equations of the paired outcomes in~(\ref{eqjoint}) are
connected via
the shared random effect vector $\widetilde{\mathbf{U}}_i$,
which is used to account for not only the correlation among the
repeated measurements from the same
subject, but also the correlation between the two outcome variables.
The subject-specific random effect is assumed to be independently
normally distributed,
that is, $\widetilde{\mathbf{U}}_i\sim
\mathcal{N}(\mathbf{0},\bolds{\Sigma}_u)$, with
variance--covariance matrix
%
\begin{equation}
\label{eqsigmau} \bolds{\Sigma}_u= \pmatrix{ \sigma_1^2
& \rho\sigma_1 \sigma_2 \vspace*{2pt}
\cr
\rho
\sigma_1 \sigma_2 & \sigma_2^2 },
\end{equation}
where $\sigma_1^2$ and $\sigma_2^2$ are two variance components, and
$\rho$ is the correlation coefficient of the random subject effects of
the paired outcomes $Y^{(1)}_{ij}$ and $Y^{(2)}_{ij}$
from the same subject measured at one visit. The correlation
structure allows us to jointly examine the two closely related clinical outcomes
in a unified modeling framework. The within-subject random errors
associated with $Y^{(1)}_{ij}$ and
$Y^{(2)}_{ij}$, namely, $\epsilon_{ij}^{(1)}$ and $\epsilon_{ij}^{(2)}$, are
assumed to follow two independent stochastic processes that could
possibly lead to serial correlation within each error series.
In general, a wide range of correlation structures could be embedded
into the errors by assuming, for example,
$\operatorname{cor}(\epsilon_{ij}^{(l)}, \epsilon_{ij'}^{(l)})=q(d(\mathbf
{s}_{ij}, \mathbf{s}_{ij'}),\bolds{\phi})$,
$l=1,2,$ where $q$ is
a correlation function taking values between $-1$ and $1$, $d$ is some
distance measure of the two position vectors
$\mathbf{s}_{ij}$ and $\mathbf{s}_{ij'}$ associated with\vspace*{1pt} $\epsilon_{ij}^{(l)}$ and $\epsilon_{ij'}^{(l)}$, respectively,
and $\bolds{\phi}$ denotes a vector of correlation parameters.
Without loss of generality, in the following discussion on model estimation
we assume simple independent errors with normal distribution $(\epsilon_{ij}^{(1)},
\epsilon_{ij}^{(2)})^T \sim
\mathcal{N}(\mathbf{0},\bolds{\Sigma}_{\epsilon})$, where
$\bolds{\Sigma}_\epsilon=\sigma_\epsilon^2 \operatorname{diag}(1,\delta^2)$ with the dispersion parameter
$\sigma_{\epsilon}^2$ and a relative scale parameter $\delta$ for
modeling heteroscedasticity of the random errors associated with the
two outcomes.
More general error processes including autoregressive moving-average
($\operatorname{ARMA}$) and continuous time autoregressive structures can be
introduced into the current modeling framework, which will be revisited
at the end of Section~\ref{secest}.

In~(\ref{eqjoint}) we assume a compound symmetry covariance for the
random subject effects,
which implies that the correlation between the two random effects
corresponding to the paired outcomes
remains constant over time. If necessary, serial correlations among
repeated measures from the same individual could be specified by the
within-subject error structure. In this study, we are primary
interested in the effect of somatic growth on blood pressure
development in children, both of
which increase with age. Hence, we opt not to explicitly introduce
temporal effects into the model.
In applications where time trajectories are of primary interest,
time-varying correlation structures can be incorporated if
constant correlation assumption is not adequate. See \citet{Morris2001}
and \citet{Dubin2005} for related discussion on the modeling of varying
correlations between random functions.

\subsection{Nonparametric smooth functions}

In the proposed semiparametric mixed model~(\ref{eqjoint}),
the group-specific bivariate smooth functions are represented by linear
combinations of some generic basis functions as follows:
\[
f^{(1)}_{g}(w,h)=\sum_{k=0}^{K}
b_{k}(w,h)\beta_{gk},\qquad f^{(2)}_{g}(w,h)=
\sum_{k=0}^{K} c_{k}(w,h)
\gamma_{gk},
\]
where $b_{0}=c_{0}=1$, $b_{k}(w,h)$ and $c_{k}(w,h), k=1,\ldots,K$,
are the basis functions of any bivariate smoother;
$\bolds{\beta}_g = (\beta_{g0}, \ldots, \beta_{gK})^T$,
$\bolds{\gamma}_g = (\gamma_{g0}, \ldots, \gamma_{gK})^T,
g=1,\ldots,G,$ are regression coefficients associated with the
nonparametric components in each group. Note that $(\beta_{10},\ldots
,\beta_{G0})$
and $(\gamma_{10},\ldots,\gamma_{G0})$ represent, respectively, the population
average of the paired outcomes in different groups.
In this research, we choose to use a thin plate spline as the
bivariate smoother [\citet{Wood2003}], which is computationally more
convenient for
modeling high-dimensional nonparametric functions.
The thin plate regression splines in a semiparametric model
can be estimated via the penalized likelihood approach [\citet{Green1994}].
The penalized log-likelihood function of model~(\ref{eqjoint}) can be
written as\vspace*{-1pt}
\[
\ell_p = \ell- \sum_{g=1}^{G}
\lambda_{g}J \bigl(f^{(1)}_{g} \bigr) - \sum
_{g=1}^{G}\varphi_{g}J
\bigl(f^{(2)}_{g} \bigr),\vspace*{-1pt}
\]
where $\ell$ is the log-likelihood function, and $J$ is the roughness
penalty functional for a bivariate twice-differentiable function $f$.
Here we write the roughness penalty $J$ of a generic function $f(x_1, x_2)$
in the following form:
\[
J(f)=\int\int_{\mathfrak{R}^2} \biggl\{ \biggl(\frac{\partial^2
f}{\partial x_1^2}
\biggr)^2 + 2 \biggl(\frac{\partial^2 f}{\partial
x_1\,
\partial x_2} \biggr)^2 + \biggl(
\frac{\partial^2
f}{\partial x_2^2} \biggr)^2 \biggr\} \,dx_1
\,dx_2.
\]

As in the unidimensional case, with the observed data, one could
express the roughness penalty as
quadratic forms of the regression coefficient vectors, that is,\vspace*{1pt}
$J(f^{(1)}_{g})=\bolds{\beta}_g^T
\bolds{\Lambda}^{\beta}_{g} \bolds{\beta}_g /2$ and
$J(f^{(2)}_{g})=\bolds{\gamma}_g^T
\bolds{\Lambda}^{\gamma}_{g} \bolds{\gamma}_g /2$, where $\bolds{\Lambda
}^{\beta}_{g}$ and
$\bolds{\Lambda}^{\gamma}_{g}$ are positive semi-definite
penalty matrices. The nonnegative smoothing parameters
$\bolds{\lambda}=(\lambda_{1}, \ldots, \lambda_{G})^T$ and
$\bolds{\varphi}=(\varphi_{1}, \ldots, \varphi_{G})^T$ control the
trade-off between goodness of fit and model smoothness. The roughness
penalties could be
further written into a more condensed form $ (\bolds{\beta}^T \bolds
{\Lambda}_{\beta}\bolds{\beta} +
\bolds{\gamma}^T \bolds{\Lambda}_{\gamma}
\bolds{\gamma} )/2$, where
$\bolds{\Lambda}_{\beta}=\operatorname{diag} (\lambda_1
\bolds{\Lambda}^{\beta}_{1}, \ldots, \lambda_G
\bolds{\Lambda}^{\beta}_{G} )$ and
$\bolds{\Lambda}_{\gamma}=\operatorname{diag} (\varphi_1
\bolds{\Lambda}^{\gamma}_{1}, \ldots, \varphi_G
\bolds{\Lambda}^{\gamma}_{G} )$ are block-diagonal penalty
matrices corresponding to the two outcomes.

With the observed values of the independent variables
$\{w_{ij}, h_{ij}\}_{1\leq j\leq n_i;
1\leq i\leq m}$, we write the smooth terms
(including the intercepts) into a matrix form
\begin{eqnarray*}
\Biggl[\sum_{g=1}^G f_g^{(1)}(w_{ij},
h_{ij})z_{ig} \Biggr]_{1\leq j\leq n_i;
1\leq i\leq
m}&=&\mathbf{X}_{\beta}
\bolds{\beta},
\\[-1.5pt]
\Biggl[\sum_{g=1}^G f_g^{(2)}(w_{ij},
h_{ij})z_{ig} \Biggr]_{1\leq j\leq n_i;
1\leq i\leq m}&=&\mathbf{X}_{\gamma}
\bolds{\gamma},\vspace*{-1pt}
\end{eqnarray*}
where\vspace*{-1pt}
\begin{eqnarray*}
\mathbf{X}_{\beta}&=& \left[\matrix{\displaystyle\mathop{{z_{i1}b_k(w_{ij},
h_{ij})}}_{0\leq k\leq
K}& \cdots& \displaystyle\mathop{z_{iG}b_k(w_{ij},h_{ij})}_{0\leq k\leq
K}
}\right]_{1\leq j\leq n_i; 1\leq i\leq
m},
\\[-1.5pt]
\bolds{\beta} &= &\bigl(\bolds{\beta}_{1}^T, \ldots,
\bolds{\beta}_{G}^T\bigr)^T,\\[-1.5pt]
\mathbf{X}_{\gamma}&=& \left[\matrix{\displaystyle\mathop{{z_{i1}c_k(w_{ij},
h_{ij})}}_{0\leq
k\leq
K}& \cdots &\displaystyle\mathop{{z_{iG}c_k(w_{ij},
h_{ij})}}_{0\leq k\leq
K} }\right]_{1\leq j\leq n_i; 1\leq i\leq
m}\vspace*{-1pt}
\end{eqnarray*}
and\vspace*{-2pt}
\[
\bolds{\gamma} = \bigl(\bolds{\gamma}_{1}^T, \ldots,
\bolds{\gamma}_{G}^T\bigr)^T.
\]
Therefore, estimation of the nonparametric
bivariate smooth functions
can be achieved through penalized estimation procedure of the
corresponding regression coefficients.
The smoothing parameters in semiparametric regression models can be
determined by, for example,
generalized cross-validation (GCV) or maximum likelihood (ML)
approaches [\citet{Wahba1985}],
among other methods. In the next section we discuss estimation of the
nonparametric components in the proposed semiparametric mixed model,
based on the ML method.

\subsection{Mixed model presentation and estimation}
\label{secest}

Let $\mathbf{Y}_1=(Y^{(1)}_{1,1}, \ldots,\break Y^{(1)}_{m,n_{m}})^{T}$,
$\mathbf{Y}_2=(Y^{(2)}_{1,1}, \ldots, Y^{(2)}_{m,n_{m}})^{T}$,
$\mathbf{Y}=(\mathbf{Y}_{1}^{T},\mathbf{Y}_{2}^{T})^{T}$
be the response variable vectors, and $N=\sum_{i=1}^m n_i$ be the
number of total observations. We denote
$\mathbf{U}_1= (U^{(1)}_1,\ldots, U^{(1)}_m )^T$,
$\mathbf{U}_2= (U^{(2)}_1,\ldots,U^{(2)}_m )^T$,
$\mathbf{U}= (\mathbf{U}_{1}^{T},\mathbf{U}_{2}^{T} )^{T}$
as the vectors of subject-specific random effects. We write
$\bolds{\epsilon}_1=(\epsilon^{(1)}_{1,1}, \ldots,
\epsilon^{(1)}_{m,n_{m}})^{T}$,
$\bolds{\epsilon}_2=(\epsilon^{(2)}_{1,1}, \ldots,
\epsilon^{(2)}_{m,n_{m}})^{T}$,
$\bolds{\epsilon}=(\bolds{\epsilon}_{1}^{T},\bolds{\epsilon}_{2}^{T})^{T}$
as the random error vectors. We denote the model matrix associated with
the parametric components in (\ref{eqjoint}) as
$\widetilde{\mathbf{T}}=\mathbf{I}_2
\otimes\mathbf{T}$ ($\mathbf{I}_n$ is the identity matrix of dimension $n$,
$\otimes$ denotes Kronecker product, and
henceforth) with corresponding parameter vector $\bolds{\psi}$,
where $\mathbf{T}= [\mathbf{t}^T_{ij} ]_{1\leq j\leq n_i; 1\leq
i\leq m}$
and $\bolds{\psi} = (\bolds{\psi}_{1}^T, \bolds{\psi}_{2}^T)^T$.
It is straightforward to set up the model
matrix $\widetilde{\mathbf{Z}}=\mathbf{I}_2
\otimes\mathbf{Z}_u$ of the random effects $\mathbf{U}$,
such that the elements of $\mathbf{Z}_u
\mathbf{U}_1$ corresponding to subject $i$ are equal to $U^{(1)}_i$,
and similarly for $\mathbf{Z}_u \mathbf{U}_2$. Then the
semiparametric mixed model for paired outcomes could be expressed in a
more condensed form as follows:
%
\begin{equation}
\mathbf{Y}=\widetilde{\mathbf{X}} \bolds{\vartheta} + \widetilde{\mathbf{Z}}
\mathbf{U} + \bolds{\epsilon}, \label{eqlmm0}
\end{equation}
where the block-diagonal matrix
$\widetilde{\mathbf{X}}= (\widetilde{\mathbf{T}},\operatorname{diag}
(\mathbf{X}_{\beta},\mathbf{X}_{\gamma} ) )$
is the model matrix of the fixed effects (including parametric and
nonparametric components), parameter vector
$\bolds{\vartheta}=(\bolds{\psi}^T,\bolds{\beta}^T,
\bolds{\gamma}^T)^T$, $\mathbf{U}\sim
\mathcal{N}(\mathbf{0},\bolds{\Sigma}_u \otimes
\mathbf{I}_m)$ is the random effects vector, and random errors
$\bolds{\epsilon}\sim
\mathcal{N}(\mathbf{0},\bolds{\Sigma}_\epsilon\otimes
\mathbf{I}_N)$. Model (\ref{eqlmm0}) can be fitted using the penalized
maximum likelihood method with roughness penalties
on the nonparametric components. Compared with the GCV approach for
choosing smoothing parameters through penalized estimation procedure,
ML-based methods are computationally more advantageous [\citet{Kohn1991}].
Furthermore, under the mixed model framework, determination of
the smoothing parameters can be naturally embedded in the model
estimation procedure [\citet{Lin1999}]. In the remainder of this section
we discuss the fitting algorithm of the proposed semiparametric mixed
model in greater detail.

As many have noted, the penalized likelihood approach has a natural
connection to the mixed effects models [\citet{Ruppert2003}, \citet{Wood2006}].
Within the mixed model framework, the nonparametric smooth terms are
treated as
regular components, with the unpenalized terms as
fixed effects and penalized terms as random effects. Because
of the unpenalized terms (e.g., the intercepts) in the
smooth components, the penalty matrices
$\bolds{\Lambda}_{\beta}$ and $\bolds{\Lambda}_{\gamma}$
are often singular; it is therefore necessary to separate the
unpenalized (fixed) and penalized (random) elements in the parameter
vectors $\bolds{\beta}$ and $\bolds{\gamma}$ so that the penalty
matrices associated with the penalized elements are of full-rank.
Specifically, we write the parameter vectors as
$\bolds{\beta}=(\bolds{\beta}^T_F,
\bolds{\beta}^T_R)^T$ and
$\bolds{\gamma}=(\bolds{\gamma}^T_F,
\bolds{\gamma}^T_R)^T$, with corresponding full-rank penalty matrices
$\mathbf{S}_{\beta}$ and $\mathbf{S}_{\gamma}$ on $\bolds{\beta}_R$ and
$\bolds{\gamma}_R$ respectively.
In this formulation, we consider
$\bolds{\beta}_F$, $\bolds{\gamma}_F$ as fixed effects, $\bolds{\beta
}_R$, $\bolds{\gamma}_R$
as random effects so that $\bolds{\beta}^T
\bolds{\Lambda}_{\beta}\bolds{\beta}=\bolds{\beta}^T_R
\mathbf{S}_{\beta}\bolds{\beta}_R$ and
$\bolds{\gamma}^T
\bolds{\Lambda}_{\gamma}\bolds{\gamma}=\bolds{\gamma}^T_R
\mathbf{S}_{\gamma}\bolds{\gamma}_R$ (note that the fixed effects
$\bolds{\beta}_F$ and $\bolds{\gamma}_F$ have zero roughness penalty).
By rewriting
the model matrices of the smooth components as
$\mathbf{X}_{\beta}= (\mathbf{X}^{\beta}_F,
\mathbf{X}^{\beta}_R )$ and
$\mathbf{X}_{\gamma}= (\mathbf{X}^{\gamma}_F,
\mathbf{X}^{\gamma}_R )$ and letting
$\bolds{\theta}=(\bolds{\psi}^T, \bolds{\beta}^T_F,
\bolds{\gamma}^T_F)^T$ be the parameters of the fixed effects,
and $\bolds{\eta}=(\mathbf{U}^T, \bolds{\beta}^T_R,
\bolds{\gamma}^T_R)^T$ be the random effects, we are able to express
the semiparametric model (\ref{eqlmm0}) as a linear mixed model (LMM):
%
\begin{equation}
\mathbf{Y}=\mathbf{X} \bolds{\theta} + \mathbf{Z} \bolds{\eta} + \bolds{
\epsilon}, \label{eqlmm1}
\end{equation}
where the block-diagonal model matrices are defined as
$\mathbf{X}\!=\!(\widetilde{\mathbf{T}},\operatorname{diag} (\mathbf
{X}^{\beta}_F,\mathbf{X}^{\gamma}_F
) )$, $\mathbf{Z}= (\widetilde{\mathbf{Z}},\operatorname{diag} (
\mathbf{X}^{\beta}_R,\mathbf{X}^{\gamma}_R  ) )$; the random
effects $\mathbf\eta\sim
\mathcal{N}(\mathbf{0},\bolds{\Sigma}_\eta)$, and random errors $\bolds
{\epsilon}\sim
\mathcal{N}(\mathbf{0},\mathbf{R})$, with $\bolds{\Sigma}_\eta=\operatorname
{diag} (\bolds{\Sigma}_u \otimes
\mathbf{I}_m, \mathbf{S}^{-1}_{\beta}, \mathbf{S}^{-1}_{\gamma}  )$,
and $\mathbf{R}=\bolds{\Sigma}_\epsilon\otimes
\mathbf{I}_N$.
From a~Bayesian perspective, under uniform and improper priors on the
fixed effects and Gaussian priors on the random effects with
variance--covariance matrix $\bolds{\Sigma}_\eta$, the penalized
likelihood estimates are simply the posterior modes. The
variance--covariance matrices $\mathbf{S}^{-1}_{\beta}$ and $\mathbf
{S}^{-1}_{\gamma}$ of the random effects $\bolds{\beta}_R$ and $\bolds
{\gamma}_R$ depend on the smoothing parameters $\bolds{\lambda}$ and
$\bolds{\varphi}$,
respectively, which can be treated as regular variance components in
the LMM.

In this LMM framework, the semiparametric mixed model can be fitted
using either ML or restricted maximum likelihood
(REML) methods [see, e.g., \citet{Lin1999}], with the smoothing
parameters treated as regular variance component parameters.
Specifically, we write $\mathbf{e}=\mathbf{Z} \bolds{\eta} + \bolds
{\epsilon}$,
and the variance component parameter vector $\bolds{\tau}=(\bolds
{\lambda}^T, \bolds{\varphi}^T,
\rho,\delta,\sigma_1^2,\sigma_2^2,\sigma_{\epsilon}^2)^T$. It then
follows that (\ref{eqlmm1}) is equivalent to
%
\begin{equation}
\mathbf{Y}=\mathbf{X} \bolds{\theta} + \mathbf{e},\qquad \mathbf{e} \sim
\mathcal{N}( \mathbf{0},\mathbf{V}), \label{eqlmm2}
\end{equation}
where $\mathbf{V}=\mathbf{Z} \bolds{\Sigma}_\eta\mathbf{Z}^T + \mathbf
{R}$ is a function of the variance components $\bolds{\tau}$. Hence,
the likelihood function given the observed response vector $\mathbf{y}$ becomes
\[
L(\bolds{\theta},\bolds{\tau})=\frac{1}{(2\pi)^N |\mathbf{V}
|^{1/2}}\exp \bigl\{(\mathbf{y}-
\mathbf{X} \bolds{\theta})^T\mathbf {V}^{-1}(\mathbf{y}-
\mathbf{X} \bolds{\theta}) \bigr\}.
\]
Model estimation of (\ref{eqlmm2}) can be achieved by maximizing the
above objective function or the REML criterion $\ell_R(\bolds{\tau
})=\log\int{L(\bolds{\theta},\bolds{\tau}) \,d\bolds{\theta}}$. The
latter has a closed-form expression
\[
\ell_R(\bolds{\tau})=-\tfrac{1}{2} \bigl\{\log|\mathbf{V}|+
\log|\mathbf {X}^T \mathbf{V}^{-1} \mathbf{X}|+(\mathbf{y}-
\mathbf{X} \widetilde {\bolds{\theta}})^T\mathbf{V}^{-1}(
\mathbf{y}-\mathbf{X} \widetilde {\bolds{\theta}}) \bigr\},
\]
where $\widetilde {\bolds{\theta}}=(\mathbf{X}^T \mathbf{V}^{-1} \mathbf
{X})^{-1}\mathbf{X}^T \mathbf{V}^{-1}\mathbf{y}$ is the generalized
least-square estimate of the fixed effects $\bolds{\theta}$ given
$\mathbf{V}$.
Statistical inferences concerning the model parameters in~(\ref
{eqlmm2}) can thus be conducted in this LMM framework.

We conclude this section with a brief comment on the correlation
structure of the random errors.
In the above discussion we have assumed an independent error structure
with variance--covariance matrix
$\mathbf{R}=\bolds{\Sigma}_\epsilon\otimes\mathbf{I}_N$ for
convenience of derivation.
In some longitudinal applications, however, such a simple error
structure may not be adequate. To accommodate
more complex error processes, we can let the variance matrix take a
more general form, for example,
$\mathbf{R}=\mathbf{R}(\delta, \sigma_{\epsilon}^2, \bolds{\phi})$,
where~$\bolds{\phi}$ is the correlation parameter vector. Serial
correlation structures
such as the often used autoregressive-moving average ($\operatorname{ARMA}$) can be
embedded into the current model framework with properly defined
variance matrix $\mathbf{R}$. If the longitudinal measurements are not
equally spaced due to design or missingness, a continuous time error
process may be adopted. For example, the continuous time autoregressive
(of order 1) structure is widely used in many applications [\citet
{Jones1993}] and it assumes
$\operatorname{cor}(\epsilon_{ij}^{(l)}, \epsilon_{ij'}^{(l)})=\phi^d,
l=1,2,$ with $d$ denoting the time interval between the two measurements
and $\phi$ being the correlation parameter of unit time interval.
See \citeauthor{Pinheiro2000} [(\citeyear{Pinheiro2000}), Section 5.3], for detailed discussion on model
specification of various error structures.

\section{Hypothesis testing}
\label{sectesting}

\subsection{Bootstrap test}
An implicit assumption of the proposed model~(\ref{eqjoint}) is that
the nonparametric
bivariate surface may interact with other independent variables. In
other words, the joint
effects of the two continuous variables may vary across different
groups. In the context of the
blood pressure study, an important scientific question is whether the
joint height--weight effects on blood pressure differ among
sex--ethnicity groups. In particular, we are primarily interested in
testing the following hypothesis in model~(\ref{eqjoint}):
%
\begin{equation}
\label{eqhypo} H_0\dvtx f^{(1)}_1=
\cdots=f^{(1)}_G,\qquad f^{(2)}_1=
\cdots=f^{(2)}_G \quad\mathrm{vs.}\quad H_A\dvtx
\mathrm{otherwise}.
\end{equation}

A likelihood ratio test could be constructed based on statistic $\Delta
=\ell(\widehat{\bolds\theta},\widehat{\bolds\tau})-\ell(\widehat{\bolds\theta}_0,\widehat{\bolds\tau}_0)$,
where $\ell(\widehat{\bolds\theta},\widehat{\bolds\tau})$ represents
the value of the log-likelihood function evaluated at the maximum
likelihood (or REML) estimates from the unrestricted model and
$\ell(\widehat{\bolds\theta}_0,\widehat{\bolds\tau}_0)$ represents
the value of log-likelihood evaluated under the null hypothesis.
\citet{Zhang2003} proposed to use a scaled $\chi^2$ distribution to
test the equivalence of two nonparametric functions in semiparametric
additive mixed models. The test they proposed considered unidimensional
smooth functions for two groups. It is much more difficult in comparing
bivariate smooth functions from multiple ($G>2$) groups, especially if
the supports of the bivariate functions are not entirely overlapping,
such as, in our application, boys and girls have different ranges of
height and weight. In the absence of theoretical development on the
sampling distribution of the likelihood-based test statistic
$\Delta$ for paired outcomes, we resort to resampling techniques for
the approximation of the empirical distribution of $\Delta$. Similar
techniques were proposed by \citet{Roca2008} for testing of
factor-by-surface interactions in a logistic generalized additive model
(GAM). We herein extend this test to paired outcome data in a
longitudinal setting.

The bootstrap testing procedure that we propose is carried out through
the following steps:
\begin{longlist}[(1)]
\item[(1)] For $j=1,\ldots,n_i$ and $i=1,\ldots,m$,
estimate (predict) the restricted mean response $\widehat{\bolds{\mu}}_{ij}$,
random subject effect $\widehat{\mathbf{U}}_i$ and random error $\hat
{\bolds{\epsilon}}_{ij}$,
from the fitted model~(\ref{eqjoint}) under the null hypothesis, where
$\widehat{\bolds{\mu}}_{ij}= (\widehat{\mu}_{ij}^{(1)},\widehat{\mu
}_{ij}^{(2)} )^T$,
$\widehat{\mathbf{U}}_i= (\widehat{U}^{(1)}_{i}, \widehat
{U}^{(2)}_{i} )^{T}$,
$\hat{\bolds{\epsilon}}_{ij}= (\hat{\epsilon}_{ij}^{(1)},\hat
{\epsilon}_{ij}^{(2)} )^T$;
\item[(2)] Draw a bootstrap sample of the random subject effects
$\widetilde{\mathbf{U}}_i\!=\! (\widetilde{U}^{(1)}_{i}, \widetilde
{U}^{(2)}_{i} )^{T}$
from $ \{\widehat{\mathbf{U}}_i \}_{1\leq i\leq m}$ with replacement;
\item[(3)] Let the bootstrap residuals be $\tilde{\epsilon}_{ij}^{(1)}=\hat
{\epsilon}_{ij}^{(1)}\varepsilon_{ij}^{(1)}$
and $\tilde{\epsilon}_{ij}^{(2)}=\hat{\epsilon}_{ij}^{(2)}\varepsilon_{ij}^{(2)}$, where
$\varepsilon_{ij}^{(1)}$ and $\varepsilon_{ij}^{(2)}$ are i.i.d.
random variables which have equal probabilities 0.5 to
be 1 or $-1$;
\item[(4)] Generate a bootstrap sample of paired responses
$\widetilde{\mathbf{Y}}_{ij}= (\widetilde{Y}^{(1)}_{ij}, \widetilde
{Y}^{(2)}_{ij} )^{T}$ by
$\widetilde{Y}^{(1)}_{ij}=\widetilde{U}^{(1)}_{i}+\widehat{\mu
}_{ij}^{(1)}+\tilde{\epsilon}_{ij}^{(1)}$ and
$\widetilde{Y}^{(2)}_{ij}=\widetilde{U}^{(2)}_{i}+\widehat{\mu
}_{ij}^{(2)}+\tilde{\epsilon}_{ij}^{(2)}$,
based on the bootstrap samples from Steps 2 and 3;
\item[(5)] Fit the joint model~(\ref{eqjoint}) to the bootstrap data
$ \{\widetilde{\mathbf{Y}}_{ij} \}_{1\leq j\leq n_i; 1\leq i\leq
m}$ under the null hypothesis and
the unrestricted model, and calculate the bootstrap test statistic~$\Delta^*$;
\item[(6)] Repeat Steps 2--5 for $b=1,\ldots,B$ times, to obtain a
bootstrap sample of the test statistic $\{\Delta^*_b\}_{1\leq b\leq B}$,
which can be used as the nominal distribution of the test statistic
under the null hypothesis.
\end{longlist}
The $p$-value of the bootstrap testing is calculated as $p = \#\{\Delta
>\Delta^*_b\}/B$.
It should be noted that in Step 3, a wild bootstrap [see, e.g., \citet
{Liu1988}, \citet{Mammen1993}] with the Rademacher distribution is used instead
of the original version, as the former has been shown to possess better
finite-sample performance [\citet{Davidson2008}].
This bootstrap procedure is partly based on the best linear unbiased
predictors (BLUPs) of the random effects,
which may underestimate the variability in the data and lead to biased
inferences [although the results are asymptotically unbiased, see \citet
{Morris2002}]. In our application, with a relatively large sample size
($m=418$ subjects with a median of 16 visits for each subject), the
bias associated with the test is likely to be negligible.

\subsection{An ad hoc likelihood ratio test}

Due to the large number of iterative fitting of complex models, the
implementation of the previously proposed resampling-based test is
computationally intensive. In this section we consider an ad hoc
likelihood ratio test (LRT) based on the asymptotic $\chi^2$
distribution, as a computationally more efficient alternative. Writing
the semiparametric mixed model~(\ref{eqjoint}) as a linear mixed model
(LMM) as in (\ref{eqlmm1}), we could construct a LRT within the LMM
framework for hypothesis~(\ref{eqhypo}). However, as noted by \citet
{Crain2004}, the asymptotic properties of LRT based on $\chi^2$
distributions [\citet{SelfLiang1987}] are not always satisfactory when
applied to penalized splines. Whereas, if no roughness penalty is added
to the smooth functions, statistical inferences (including significance
tests) will have more reasonable behaviors for unpenalized models [\citet
{Wood2006}, page 195], at the price of overfitting. Additionally, LRT
for fixed effects based on standard $\chi^2_\nu$ distributions (with
the degrees of freedom $\nu$ being the difference of the numbers of
parameters between the null and unrestricted) tends to be more
``anticonservative'' [\citet{Pinheiro2000}, see discussions on pages
87--88]. To alleviate, we adopt a mixture of $\chi^2_\nu$ and $\chi^2_{\nu+1}$ as the reference distribution suggested by \citet
{Stram1994}, the empirical performance of which is studied in
Section~\ref{secsim}.
The adjusted LRT is conducted through the following steps:
\begin{longlist}[(1)]
\item[(1)] Fit model~(\ref{eqjoint}) using penalized splines based on ML to
obtain the effective degrees of freedom (EDF) for the penalized spline
estimates;
\item[(2)] Refit (\ref{eqjoint}) under the null and unrestricted models,
respectively, by fixing the degrees of freedom to be (approximately)
the estimated EDF from Step 1 using the unpenalized splines;
\item[(3)] Calculate the $p$-value based on 2 times the log-likelihood ratio
from Step 2, using $\frac{1}{2}\chi^2_\nu+\frac{1}{2}\chi^2_{\nu+1}$ as
the nominal distribution.
\end{longlist}

This alternative LRT procedure significantly reduces the computational
burden of the aforementioned inference. The resampling-based testing
procedure, on the other hand, is methodologically better grounded and
is more likely to have superior finite-sample performance. Nonetheless,
despite the ad hoc nature of the LRT, it might be able to provide
quick testing results with reasonable accuracy. The justification of
the LRT is entirely empirical. To that end, we conduct a Monte Carlo
study to assess the operational characteristics of the LRT (see
Section~\ref{secsim}). In practice, we recommend the use of the
resampling-based testing procedure whenever computing resources are
available. In the absence of adequate computing power, the LRT may
provide a reasonable relief, but the test results should be interpreted
with caution, especially for borderline cases.

\section{Monte Carlo study}
\label{secsim}

To assess the performance of the likelihood ratio test on the
significance of factor-by-surface interaction in the semiparametric
mixed model, we conduct a Monte Carlo study. Simulation results are
presented in this section.

The simulated data are generated from two nonlinear bivariate test
functions $f_1$ and $f_2$, defined on $[0,1]\times[0,1]$:
$f_{1}(x,t)=5x^2+\log(0.5t+1)+t+3t^{0.5x +1}$, $f_{2}(x,t)=1.5\sqrt {x}+1.5t^3 + 2.25x e^{t}$.
The two correlated outcome variables $ (Y^{(1)}_{ij},
Y^{(2)}_{ij} )$ are generated for $i=1,\ldots,m$ and $j=1,\ldots,n$ from
\[
\cases{ Y^{(1)}_{ij} = U^{(1)}_{i} +
\beta_0 +z_i \beta_1 +\bar
{f}_1(w_{ij},h_{ij})+ \epsilon^{(1)}_{ij},
\vspace*{2pt}
\cr
Y^{(2)}_{ij} = U^{(2)}_{i}
+ \gamma_0 +z_i\gamma_1 +\bar
{f}_2(w_{ij},h_{ij})+\epsilon^{(2)}_{ij},}
\]
where $ (U^{(1)}_{ij}, U^{(2)}_{ij} )^T\sim\mathcal{N}(\mathbf
{0},\bolds{\Sigma}_u)$ as in (\ref{eqsigmau}) with $\sigma_1=2, \sigma_2=3$, and $\rho=0.5$; $ (\epsilon_{ij}^{(1)},
\epsilon_{ij}^{(2)} )^T \sim
\mathcal{N}(\mathbf{0},\sigma_\epsilon^2 \operatorname{diag}(1,\delta^2))$, with $\sigma_\epsilon=2$ and $\delta=0.8$;
the first $m/2$ subjects are labeled as group 1 and the remaining
belonged to group 2, $z_i$ is the group indicator variable; the true
bivariate covariate effects of $(w, h)$ are assumed to be homogeneous
across groups with functional forms of $\bar{f}_1$ and $\bar{f}_2$
($\bar{f}$ denotes corresponding smooth functions centered at the
observed covariate values) for the two outcomes respectively; but the
two groups have different intercepts with $\beta_0=10, \beta_1=2, \gamma_0=15$, and $\gamma_1=4$.

We then conducted the proposed likelihood ratio tests based on the
unpenalized spline estimates and mixture $\chi^2$ distribution. We
examined two levels of $m=50, 100$ and two levels of $n=20, 40$. The
size of the test in each scenario was based on 500 replications and
summarized in Table~\ref{tabsim}, which was observed to be very close
to its nominal level 0.05 in each case.

%
\begin{table}
\tablewidth=250pt
\caption{Simulation results for likelihood ratio tests,
with nominal level 0.05. The results were based on 500 replications}\label{tabsim}
\begin{tabular*}{250pt}{@{\extracolsep{\fill}}lccc@{}}
\hline
$\bolds{m}$ & $\bolds{n}$ & \textbf{Distribution} & \textbf{Size} \\
\hline
\phantom{0}{50} & {20} & $\chi^2_\nu$ & 0.060 \\[2pt]
& & $\chi^2_{\nu+1}$ & 0.048 \\[2pt]
& & $\frac{1}{2}\chi^2_\nu+\frac{1}{2}\chi^2_{\nu+1}$ & 0.052 \\[4pt]
\phantom{0}{50} &{40} & $\chi^2_\nu$ & 0.052 \\[2pt]
& & $\chi^2_{\nu+1}$ & 0.036 \\[2pt]
& & $\frac{1}{2}\chi^2_\nu+\frac{1}{2}\chi^2_{\nu+1}$ & 0.042 \\[4pt]
{100} & {20} & $\chi^2_\nu$ & 0.062 \\[2pt]
& & $\chi^2_{\nu+1}$ & 0.044 \\[2pt]
& & $\frac{1}{2}\chi^2_\nu+\frac{1}{2}\chi^2_{\nu+1}$ & 0.054 \\[4pt]
{100} & {40} & $\chi^2_\nu$ & 0.056 \\[2pt]
& & $\chi^2_{\nu+1}$ & 0.042 \\[2pt]
& & $\frac{1}{2}\chi^2_\nu+\frac{1}{2}\chi^2_{\nu+1}$ & 0.046 \\
\hline
\end{tabular*}
\end{table}

\section{Analysis of blood pressure data}
\label{seccase}

\subsection{A childhood blood pressure development study}

\begin{figure}

\includegraphics{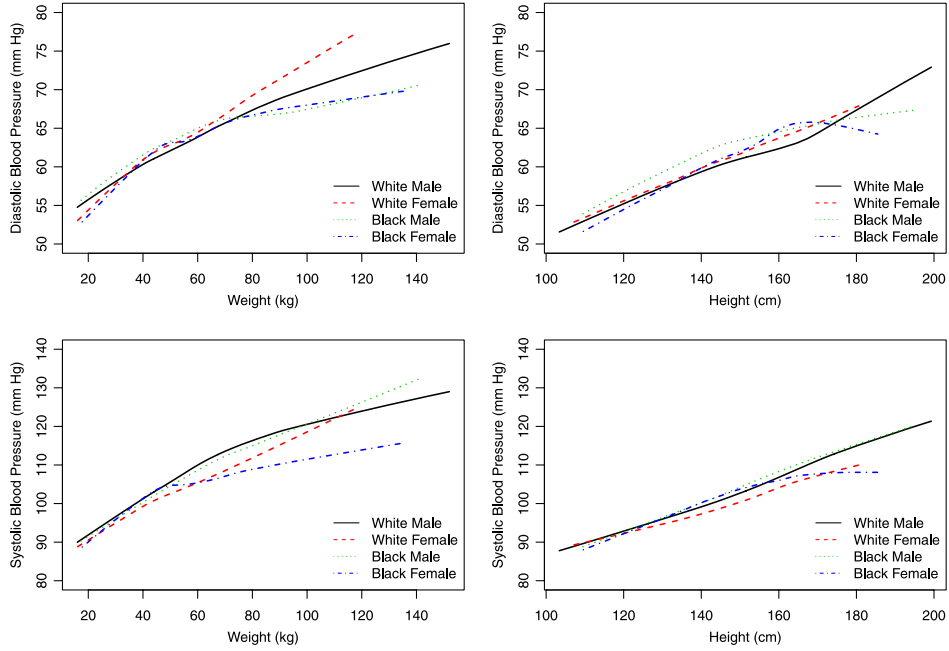}

\caption{Marginal effects of weight and height on blood pressure using
a LOWESS smoother.}
\label{figmarginal}
\end{figure}

Children from local schools were recruited for participation in a
prospective cohort study. Those with known cardiovascular disease,
hypertension, kidney disease, and those on blood pressure altering
medications were excluded. Blood pressure, height, weight and heart
rate are measured semi-annually. The study is currently ongoing. In
this analysis, we use a subset of $m=418$ children that have at least
ten ($\geq$10) semi-annual assessments. The data set includes 154
white boys (sex--ethnicity group~1, or group~1 for short and
henceforth), 136 white girls (group~2), 70 black boys (group~3) and 58
black girls (group 4). Figure~\ref{figmarginal} shows\vadjust{\goodbreak} the marginal
effects of height and weight on systolic and diastolic blood pressure
using scatterplot smoothing [\citet{Cleveland1979}], for each of the
sex--ethnicity combinations. Examining the estimated marginal effects,
we see clear indications of nonlinear weight effect and different
patterns of height and weight effects in boys and girls of different ethnicity.

To accommodate these data features, we perform an analysis using the
proposed semiparametric mixed model that simultaneously assesses the
height and weight effects on systolic and diastolic blood pressure. We
present bivariate effect surfaces in colored contour plots for the
examination of the joint height--weight effects. We also perform
resampling-based and LRT-based inferences for the detection of possible
gender and ethnicity differences in these bivariate surfaces.

%

\subsection{Model specification}

The following model is used to examine the joint height--weight effects
on diastolic and systolic blood pressure in children:

%
\begin{equation}
\label{eqbpjoint} \cases{ \displaystyle \operatorname{DBP}_{ij} =
U^d_{i} + p_{ij}\psi_d + \sum
_{g=1}^{4} f^{d}_{g}(w_{ij},
h_{ij})z_{ig}+\epsilon_{ij}^{d},
\vspace*{2pt}
\cr
\displaystyle \operatorname{SBP}_{ij} =
U^s_{i} + p_{ij} \psi_s + \sum
_{g=1}^{4} f^{s}_{g}(w_{ij},
h_{ij})z_{ig}+ \epsilon_{ij}^{s},}
\end{equation}
where $\operatorname{DBP}_{ij}$, $\operatorname{SBP}_{ij}$,
$p_{ij}$, $w_{ij}$ and $h_{ij}$, respectively, represent the diastolic
and systolic blood pressure, heart rate (or pulse, which is used as a
surrogate marker for cardiac output), weight and height measured from
the $i$th subject at the $j$th visit;
$U^d_{i}$ and $U^s_{i}$ are the random subject effects;
$\psi_d$ and $\psi_s$ are the regression coefficient parameters of
heart rate on diastolic and systolic blood pressure respectively;
$f^{d}_{g}$ and $f^{s}_{g}$ are the unknown bivariate smooth
functions to depict the joint weight and height effects on
diastolic and systolic blood pressure, respectively, in the four
sex--ethnicity groups ($g=1,\dots,4$);
and $z_{ig}$ is the corresponding group indicator.
Note that the intercept terms $(\beta_{10},\ldots,\beta_{40})$
and $(\gamma_{10},\ldots,\gamma_{40})$ representing, respectively, the
population
average diastolic and systolic blood pressure in different
sex--ethnicity groups are
absorbed into the corresponding group-specific smooth components.
The effects of heart rate were found to be linear in preliminary analyses.
Hence, they are included in the model as linear components for ease of
clinical interpretation.
The random subject effects are assumed to have the same distribution as
specified in model~(\ref{eqjoint}).

Since the outcomes are measured repeatedly for each subject during the
follow-up, possible serial correlations may exist.
According to the study protocol, enrolled subjects were asked to return
every six months for measurements after the baseline screening.
However, the longitudinal data collection was not exactly evenly spaced
due to delayed or even missed clinic visits.
To accommodate, we incorporate a continuous time autoregressive
structure into the within-subject errors. To be more exact, we assume
that $\operatorname{cor}(\epsilon_{ij}^{d}, \epsilon_{ij'}^{d}) =
\operatorname{cor}(\epsilon_{ij}^{s}, \epsilon_{ij'}^{s}) =
\phi^{|\mathit{age}_{ij}-\mathit{age}_{ij'}|}$,
for $1\leq j\neq j' \leq n_i$, $i=1,\ldots,m$, where $\mathit{age}_{ij}$
denotes the age (in years) of subject $i$ at the $j$th visit and
$\phi$ is the autocorrelation parameter of unit time interval.

The core model fitting procedure is based on the \texttt{gamm}
(generalized additive mixed model) routine in \textsf{R}
package mgcv [\citet{mgcv}]. We have made necessary extensions to
accommodate the complex model features (e.g., correlation structure of
paired longitudinal outcomes) and visualization of the results. The
confidence intervals of the model parameters are derived from the
observed information matrix in the LMM framework. The estimated
bivariate smooth functions of weight and height in each sex--ethnicity
group are presented and compared using colored image plots and contour
lines. A~detailed description of the model-fitting algorithms can be found in Section A of
the supplementary materials [\citet{LiuTu2012S}], together with model
diagnostics for model assumption verification and goodness-of-fit
assessment in Section B.

\subsection{Analytical results}

From the REML estimates (which are very close to the ML estimates) of
the semiparametric mixed model (\ref{eqbpjoint})
based on 6867 pairs of blood pressure assessments from 418 subjects,
we note a substantial correlation ($\widehat{\rho}=0.52$
with 95\% confidence interval (CI) $[0.42, 0.61]$) between the
diastolic and systolic blood pressure within the same subject. Systolic
blood pressure has slightly greater variability ($\widehat{\sigma
}_2=5.29;$ 95\% CI: $[4.88, 5.73]$) than diastolic measurements
($\widehat{\sigma}_1=4.57;$ 95\% CI: $[4.18, 4.99]$), but the
difference is not statistically significant. However, the random error
associated with systolic blood pressure (within the same subject) has a
significantly smaller variance, as reflected by the magnitude and
corresponding confidence interval of the scaling parameter ($\widehat
{\delta}=0.87;$ 95\% CI: $[ 0.85, 0.90]$). Such an observation is
consistent with the previously published data on pediatric blood
pressure measurements [\citet{Falkner2006}], and it may in part reflect
the difficulty in clearly pinpointing the start of the fifth Korotkoff
sound in diastolic measurement [\citet{Pickering2005}]. The estimated
variance of the random error is $\widehat{\sigma}_{\epsilon}=7.39$;
95\% CI: $[7.26, 7.53]$. We also detected slight autocorrelation in
within-subject errors, with $\widehat{\phi}=0.014;$ 95\% CI: $[0.010,
0.020]$. The estimates of the average diastolic
and systolic blood pressure in different gender and ethnicity groups in
model~(\ref{eqbpjoint}) are listed in Table~\ref{tabpar}.
Heart rate has a negative effect on the diastolic blood pressure with
$\psi_d=-0.04$ [standard deviation (SD)${} = {}$0.01], whereas it is
positively associated with
systolic blood pressure $\psi_s=0.07$ (SD${} = {}$0.01). The finding is not
surprising because heart rate directly reflects cardiac output, which
typically increases with pulse pressure (systolic minus diastolic blood
pressure). With pulse pressure relating systolic positively and
diastolic negatively,
one would expect systolic blood pressure to increase with
heart rate and diastolic blood pressure to decrease with heart rate.\vadjust{\goodbreak}

%
\begin{table}
\caption{Estimates of the parameters of the
semiparametric joint blood pressure model with 95\% lower (confidence)
bound (LB)
and upper (confidence) bound (UB)}\label{tabpar}
\begin{tabular*}{\textwidth}{@{\extracolsep{\fill}}lccccc@{}}
\hline
\textbf{Parameter} & \textbf{Sex--ethnicity group} & \textbf{Estimate} & \textbf{Std. Dev.} & \textbf{95\% LB} &
\textbf{95\% UB} \\
\hline
$\beta_{10}$ & White male & \phantom{0}$64.99$ & $0.98$ & 63.03 & \phantom{0}66.95 \\
$\beta_{20}$ & White female & \phantom{0}$66.02$ & $0.99$ & 64.04 & \phantom{0}68.00 \\
$\beta_{30}$ & Black male & \phantom{0}$65.98$ & $1.09$ & 63.80 & \phantom{0}68.16 \\
$\beta_{40}$ & Black female & \phantom{0}$65.76$ & $1.21$ & 63.34 & \phantom{0}68.18 \\
$\gamma_{10}$ & White male & $100.41$ & $0.92$ & 98.57 & 102.25 \\
$\gamma_{20}$ & White female & \phantom{0}$97.51$ & $0.93$ & 95.65 & \phantom{0}99.37 \\
$\gamma_{30}$ & Black male & \phantom{0}$99.93$ & $1.05$ & 97.83 & 102.03 \\
$\gamma_{40}$ & Black female & \phantom{0}$98.27$ & $1.17$ & 95.93 & 100.61 \\
\hline
\end{tabular*}
\end{table}

\begin{figure}

\includegraphics{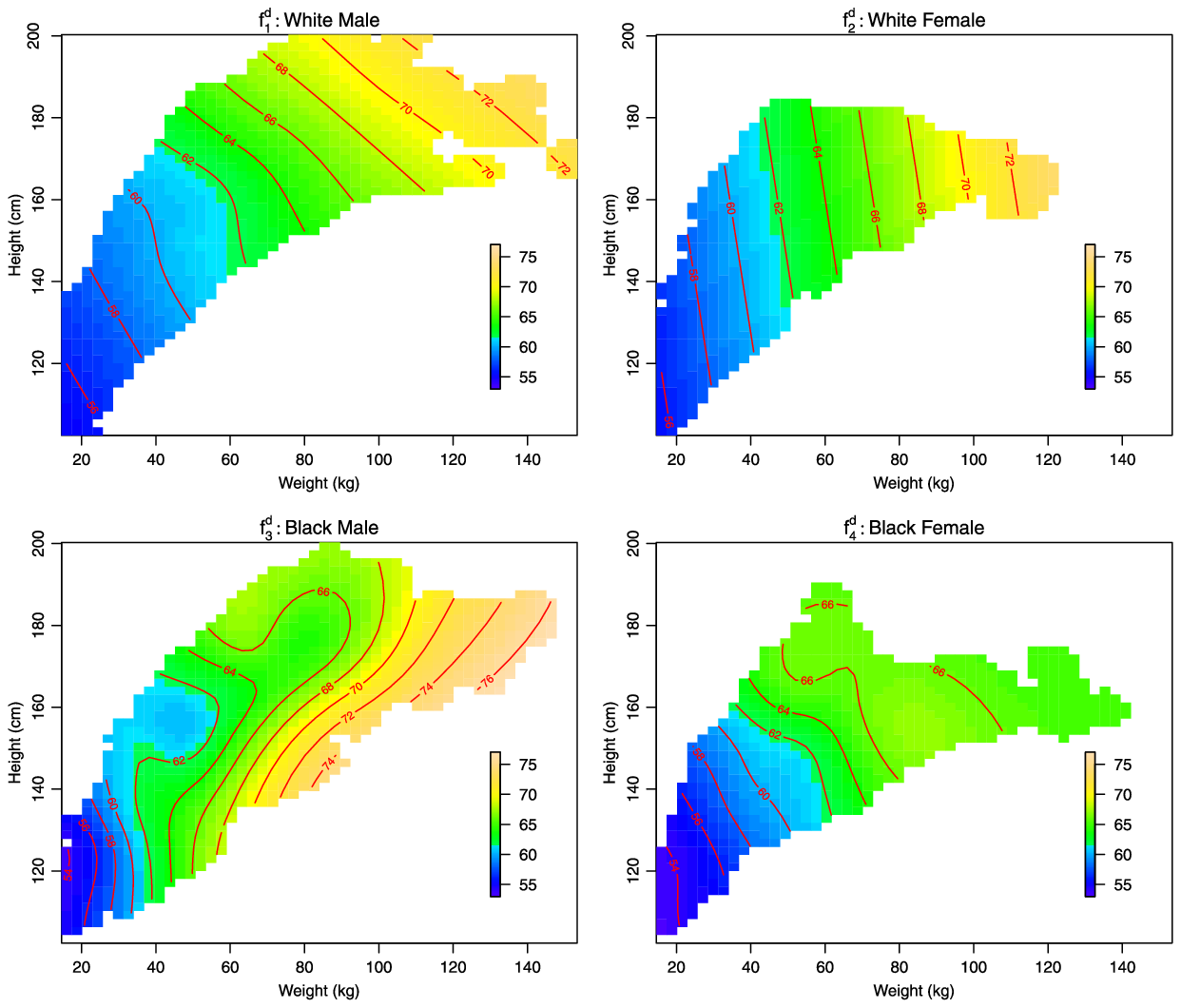}

\caption{Bivariate smooth function estimates of the joint
height--weight effects on diastolic blood pressure [defined in
equation (\protect\ref{eqbpjoint})]
in boys and girls of different ethnicity.}
\label{figdbpgroup}
\end{figure}

\begin{figure}

\includegraphics{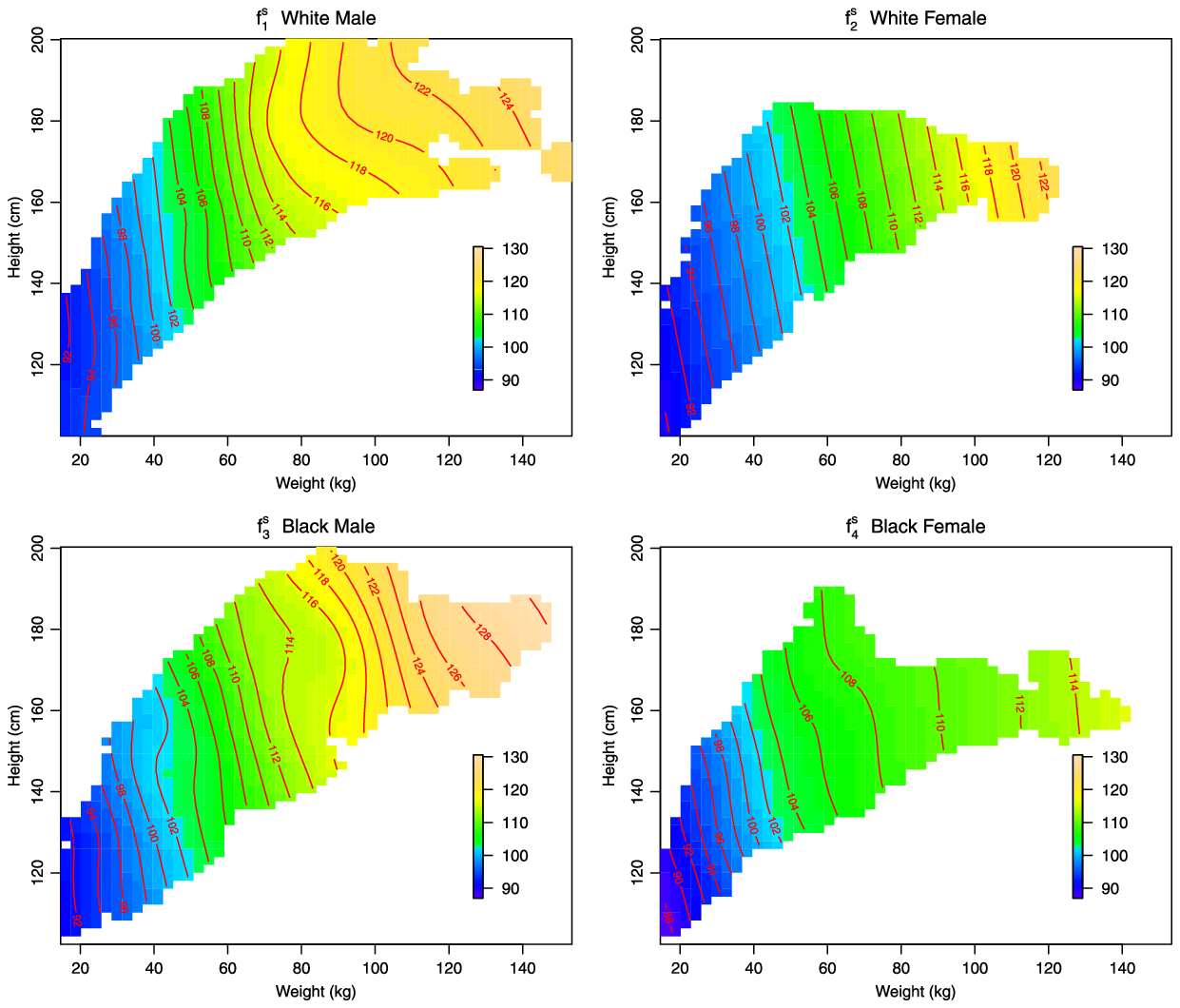}

\caption{Bivariate smooth function estimates of the joint
height--weight effects on systolic blood pressure [defined in
equation~(\protect\ref{eqbpjoint})] in boys and girls of different ethnicity.}
\label{figsbpgroup}
\end{figure}

The estimated bivariate smooth functions of weight and height in
the four sex--ethnicity groups are plotted in Figures~\ref
{figdbpgroup} and
\ref{figsbpgroup}.

%

%

\begin{figure}

\includegraphics{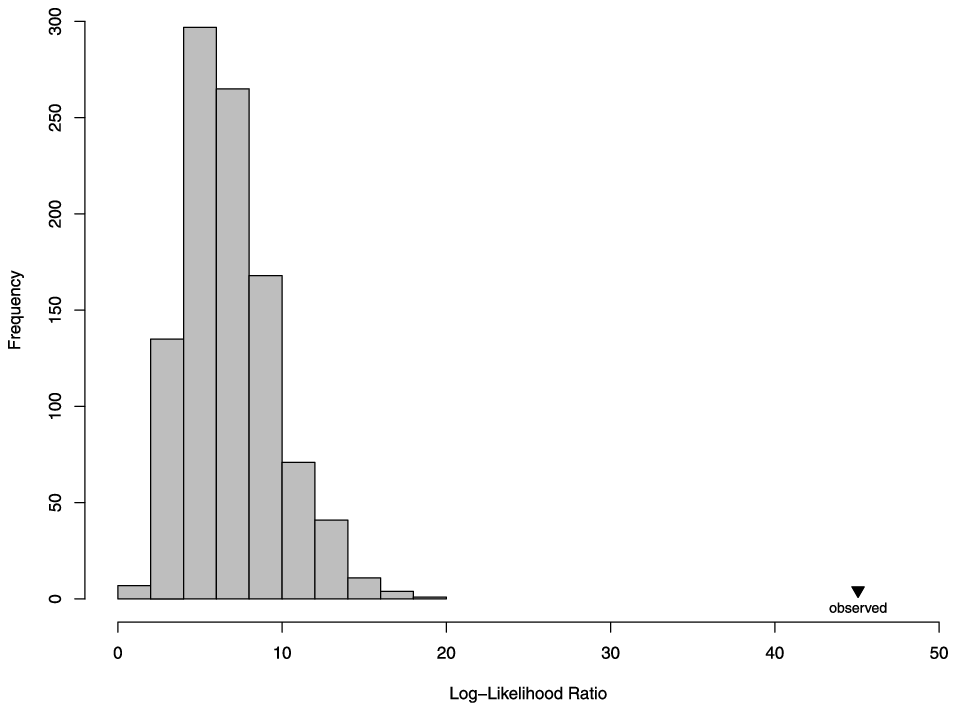}

\caption{Empirical distribution of log-likelihood ratio based on bootstrap
technique with $B=1000$ replications.}
\label{figboot}
\end{figure}

Both the adjusted likelihood ratio (LR) test [$2\log LR=217.6$ with
reference distribution
$(\chi^2_{84}+\chi^2_{85})/2$, $p<0.001$] and the bootstrap test
(Figure~\ref{figboot} with $B=1000$ replications)
suggest significantly different joint height--weight influences on blood
pressure across the four sex and ethnicity groups.
Aside from the significant test results of the bivariate height--weight
effect surfaces, the most interesting observation from this analysis is
the apparently different shape of these bivariate functions: (1) While
blood pressure generally increases with weight as well as height,
weight clearly has a much greater overall influence on blood pressure.
In fact, at a given weight level, the height effects are often minimal,
as indicated by the (nearly) vertical contour lines. (2) Among the
heavier boys (those with weight greater than 120~kg, e.g.),
blacks appear to have higher systolic blood pressure than whites. The
reverse is true for girls. From the fitted effect surfaces, we see that
heavier white girls appear to have higher systolic blood pressure than
their black counterparts. (3) For diastolic blood pressure, while
weight is still the dominant influence, height does have an effect. The
more intriguing observation is perhaps the clear difference between
black and white boys. For example, when weight is about 80~kg, taller
black boys have lower diastolic blood pressure. While one would be
attempted to attribute this to the lower corresponding BMI values, the
opposite is true for white boys. These more complex pictures of height
and weight influences on blood pressure point to possible \mbox{existence} of
distinct physiology of blood pressure development in black and white
children.

%

\section{Discussion}
\label{secdiscussion}

In summary, we have presented a joint model for paired outcomes with
bivariate effect surfaces of two continuous independent variables. This
model extends previous work by accommodating longitudinally measured
outcome pairs, as well as bivariate covariate effect surfaces. With the
introduction of factor-by-surface interactions, it also allows for the
incorporation of group-specific surfaces (i.e., group-height--weight
interactions). For implementation, we have developed necessary
computational procedures. Resampling-based and LRT-based inferences
concerning the group-specific bivariate effects are discussed.
Simulation study indicates adequate performances in the proposed
adjusted LRT procedure.

Using the proposed method, we examined the influences of height and
weight on blood pressure in children undergoing the pubertal growth
process. For adults, there is a generally accepted notion that body
weight has a predominant influence on blood pressure. For children that
undergo the pubertal growth process, height and weight are known to
increase concurrently with age, and both height and weight are
positively related to blood pressure. Few studies have directly
examined the relative contributions of height and weight, partly due to
the lack of appropriate analytical tools to discern these simultaneous
effects. With the newly developed statistical method, we examined the
influences of height and weight on longitudinally measured blood
pressure. We found that in children, just like in adults, weight tended
to have a noticeably stronger impact on blood pressure, even during a
period of vigorous linear growth. The study finding provides direct
evidence that adiposity, as reflected by weight, is the primary driver
of the blood pressure development in children. The finding is
consistent with the latest pediatric literature on the connection
between adiposity and blood pressure. Clinically, our finding
highlights the importance of weight management in overweight and obese
children: excessive weight gain could significantly increase the
hypertension risk in children [\citet{Tu2011}, \citet{Falkner2011}].
Mechanistically, weight-mediated blood pressure elevation in young and
healthy children points to the need for studies focusing on including
adipose-derived hormones.
Animal and human studies suggested one of such hormones, leptin, could
act upon the sympathetic nervous system (SNS), which contributes to the
elevation of blood pressure. More recently, investigators have proposed
mouse models describing new signaling pathways in the pathogenesis of
obesity hypertension through leptin [\citet{Purkayastha2011}, \citet{Humphreys2011}].
Interestingly, in a separate investigation, our
research team has also discovered dramatically increased leptin levels
and heart rate (as an indicator for a more activated SNS) in overweight
and obese children, corresponding to the elevated blood pressure [\citet
{Tu2011}]. The findings from the current study certainly gives more
credence to this hypothesized pathway between adiposity and blood pressure.
We are currently investigating the possibility of adiposity acting upon
blood pressure through alternative pathways, such as the
renin-angiotensin-aldosterone system.

A limitation of this research is that we only included children who
had ten or more longitudinal assessments in this methodological
development, following a condition stipulated by our current data use
agreement. While such a restriction could potentially limit the
generalizability of the study finding, we note that the number of
assessments is not related to the study subjects' behaviors but to the
timing of their school's participation into the original study. As a
result, children who had fewer assessments (and thus were excluded from
the current analysis) are unlikely to be systematically different from
those who contributed more observations. Notwithstanding this
limitation, our research does provide an important analytical tool that
may significantly enhance mechanistic and epidemiologic investigations
concerning blood pressure development in children.

\section*{Acknowledgments}
The authors wish to thank the Regenstrief Institute for its
support of this research.
We also thank the Editor, Associate Editor and two reviewers whose comments
resulted in great improvement in the manuscript.

\begin{supplement}\label{supp} 
\stitle{Detailed model-fitting algorithm and model diagnostics\\}
\slink[doi]{10.1214/12-AOAS567SUPP} 
\slink[url]{http://lib.stat.cmu.edu/aoas/567/supplement.pdf}
\sdatatype{.pdf}
\sdescription{We provide the computational details of the model-fitting
algorithm with sample R code and an R function to
visualize the predicted bivariate surfaces. Some model diagnostics
plots are also provided.}
\end{supplement}


%

\printaddresses


\begin{thebibliography}{57}

\bibitem[\protect\citeauthoryear{Anderson and Whitaker}{2009}]{Anderson2009}
%
\begin{barticle}[author]
\bauthor{\bsnm{Anderson},~\bfnm{S.~E.}\binits{S.~E.}} \AND
\bauthor{\bsnm{Whitaker},~\bfnm{R.~C.}\binits{R.~C.}}
(\byear{2009}).
\btitle{Prevalence of obesity among US preschool children in different racial
and ethnic groups}.
\bjournal{Archives of Pediatrics \& Adolescent Medicine}
\bvolume{163}
\bpages{344--348}.
\bptok{imsref}%
\end{barticle}
%
\endbibitem

\bibitem[\protect\citeauthoryear{Baker, Olsen and Sorensen}{2007}]{Baker2007}
%
\begin{barticle}[author]
\bauthor{\bsnm{Baker},~\bfnm{J.~L.}\binits{J.~L.}},
\bauthor{\bsnm{Olsen},~\bfnm{L.~W.}\binits{L.~W.}} \AND
\bauthor{\bsnm{Sorensen},~\bfnm{TIA}\binits{T.}}
(\byear{2007}).
\btitle{Childhood body-mass index and the risk of coronary heart
disease in
adulthood}.
\bjournal{New England Journal of Medicine}
\bvolume{357}
\bpages{2329--2337}.
\bptok{imsref}%
\end{barticle}
%
\endbibitem

\bibitem[\protect\citeauthoryear{Brady et~al.}{2010}]{Brady2010}
%
\begin{barticle}[pbm]
\bauthor{\bsnm{Brady},~\bfnm{Tammy~M.}\binits{T.~M.}},
\bauthor{\bsnm{Fivush},~\bfnm{Barbara}\binits{B.}},
\bauthor{\bsnm{Parekh},~\bfnm{Rulan~S.}\binits{R.~S.}} \AND
\bauthor{\bsnm{Flynn},~\bfnm{Joseph~T.}\binits{J.~T.}}
(\byear{2010}).
\btitle{Racial differences among children with primary hypertension}.
\bjournal{Pediatrics}
\bvolume{126}
\bpages{931--937}.
\bid{doi={10.1542/peds.2009-2972}, issn={1098-4275}, pii={peds.2009-2972},
pmid={20956429}}
\bptok{imsref}%
\end{barticle}
%
\endbibitem

\bibitem[\protect\citeauthoryear{Brezger, Fahrmeir and
Hennerfeind}{2007}]{Brezger2007}
%
\begin{barticle}[mr]
\bauthor{\bsnm{Brezger},~\bfnm{A.}\binits{A.}},
\bauthor{\bsnm{Fahrmeir},~\bfnm{L.}\binits{L.}} \AND
\bauthor{\bsnm{Hennerfeind},~\bfnm{A.}\binits{A.}}
(\byear{2007}).
\btitle{Adaptive {G}aussian {M}arkov random fields with applications in human
brain mapping}.
\bjournal{J. Roy. Statist. Soc. Ser. C}
\bvolume{56}
\bpages{327--345}.
\bid{doi={10.1111/j.1467-9876.2007.00580.x}, issn={0035-9254}, mr={2370993}}
\bptok{imsref}%
\end{barticle}
%
\endbibitem

\bibitem[\protect\citeauthoryear{Cleveland}{1979}]{Cleveland1979}
%
\begin{barticle}[mr]
\bauthor{\bsnm{Cleveland},~\bfnm{William~S.}\binits{W.~S.}}
(\byear{1979}).
\btitle{Robust locally weighted regression and smoothing scatterplots}.
\bjournal{J. Amer. Statist. Assoc.}
\bvolume{74}
\bpages{829--836}.
\bid{issn={0003-1291}, mr={0556476}}
\bptok{imsref}%
\end{barticle}
%
\endbibitem

\bibitem[\protect\citeauthoryear{Crainiceanu, Diggle and
Rowlingson}{2008}]{Crain2008}
%
\begin{barticle}[mr]
\bauthor{\bsnm{Crainiceanu},~\bfnm{Ciprian~M.}\binits{C.~M.}},
\bauthor{\bsnm{Diggle},~\bfnm{Peter~J.}\binits{P.~J.}} \AND
\bauthor{\bsnm{Rowlingson},~\bfnm{Barry}\binits{B.}}
(\byear{2008}).
\btitle{Bivariate binomial spatial modeling of \textit{{L}oa loa}
prevalence in
tropical {A}frica}.
\bjournal{J. Amer. Statist. Assoc.}
\bvolume{103}
\bpages{21--37}.
\bid{doi={10.1198/016214507000001409}, issn={0162-1459}, mr={2420211}}
\bptnote{check related}%
\bptok{imsref}%
\end{barticle}
%
\endbibitem

\bibitem[\protect\citeauthoryear{Crainiceanu and Ruppert}{2004}]{Crain2004}
%
\begin{barticle}[mr]
\bauthor{\bsnm{Crainiceanu},~\bfnm{Ciprian~M.}\binits{C.~M.}} \AND
\bauthor{\bsnm{Ruppert},~\bfnm{David}\binits{D.}}
(\byear{2004}).
\btitle{Likelihood ratio tests in linear mixed models with one variance
component}.
\bjournal{J. R. Stat. Soc. Ser. B Stat. Methodol.}
\bvolume{66}
\bpages{165--185}.
\bid{doi={10.1111/j.1467-9868.2004.00438.x}, issn={1369-7412}, mr={2035765}}
\bptok{imsref}%
\end{barticle}
%
\endbibitem

\bibitem[\protect\citeauthoryear{Davidson and Flachaire}{2008}]{Davidson2008}
%
\begin{barticle}[mr]
\bauthor{\bsnm{Davidson},~\bfnm{Russell}\binits{R.}} \AND
\bauthor{\bsnm{Flachaire},~\bfnm{Emmanuel}\binits{E.}}
(\byear{2008}).
\btitle{The wild bootstrap, tamed at last}.
\bjournal{J. Econometrics}
\bvolume{146}
\bpages{162--169}.
\bid{doi={10.1016/j.jeconom.2008.08.003}, issn={0304-4076}, mr={2459651}}
\bptok{imsref}%
\end{barticle}
%
\endbibitem

\bibitem[\protect\citeauthoryear{Davy and Hall}{2004}]{Davy2004}
%
\begin{barticle}[author]
\bauthor{\bsnm{Davy},~\bfnm{K.~P.}\binits{K.~P.}} \AND
\bauthor{\bsnm{Hall},~\bfnm{J.~E.}\binits{J.~E.}}
(\byear{2004}).
\btitle{Obesity and hypertension: Two epidemics or one?}
\bjournal{American Journal of Physiology---Regulatory Integrative and Comparative
Physiology}
\bvolume{286}
\bpages{R803--R813}.
\bptok{imsref}%
\end{barticle}
%
\endbibitem

\bibitem[\protect\citeauthoryear{Dean, Nathoo and Nielsen}{2007}]{Dean2007}
%
\begin{barticle}[mr]
\bauthor{\bsnm{Dean},~\bfnm{C.~B.}\binits{C.~B.}},
\bauthor{\bsnm{Nathoo},~\bfnm{F.}\binits{F.}} \AND
\bauthor{\bsnm{Nielsen},~\bfnm{J.~D.}\binits{J.~D.}}
(\byear{2007}).
\btitle{Spatial and mixture models for recurrent event processes}.
\bjournal{Environmetrics}
\bvolume{18}
\bpages{713--725}.
\bid{doi={10.1002/env.870}, issn={1180-4009}, mr={2408940}}
\bptok{imsref}%
\end{barticle}
%
\endbibitem

\bibitem[\protect\citeauthoryear{Dubin and M{\"u}ller}{2005}]{Dubin2005}
%
\begin{barticle}[mr]
\bauthor{\bsnm{Dubin},~\bfnm{Joel~A.}\binits{J.~A.}} \AND
\bauthor{\bsnm{M{\"u}ller},~\bfnm{Hans-Georg}\binits{H.-G.}}
(\byear{2005}).
\btitle{Dynamical correlation for multivariate longitudinal data}.
\bjournal{J. Amer. Statist. Assoc.}
\bvolume{100}
\bpages{872--881}.
\bid{doi={10.1198/016214504000001989}, issn={0162-1459}, mr={2201015}}
\bptok{imsref}%
\end{barticle}
%
\endbibitem

\bibitem[\protect\citeauthoryear{Falkner}{2010}]{Falkner2010}
%
\begin{barticle}[pbm]
\bauthor{\bsnm{Falkner},~\bfnm{Bonita}\binits{B.}}
(\byear{2010}).
\btitle{Hypertension in children and adolescents: Epidemiology and natural
history}.
\bjournal{Pediatr. Nephrol.}
\bvolume{25}
\bpages{1219--1224}.
\bid{doi={10.1007/s00467-009-1200-3}, issn={1432-198X}, pmcid={2874036},
pmid={19421783}}
\bptok{imsref}%
\end{barticle}
%
\endbibitem

\bibitem[\protect\citeauthoryear{Falkner and Gidding}{2011}]{Falkner2011}
%
\begin{barticle}[pbm]
\bauthor{\bsnm{Falkner},~\bfnm{Bonita}\binits{B.}} \AND
\bauthor{\bsnm{Gidding},~\bfnm{Samuel}\binits{S.}}
(\byear{2011}).
\btitle{Childhood obesity and blood pressure: Back to the future?}
\bjournal{Hypertension}
\bvolume{58}
\bpages{754--755}.
\bid{doi={10.1161/HYPERTENSIONAHA.111.180430}, issn={1524-4563},
mid={NIHMS330276}, pii={HYPERTENSIONAHA.111.180430}, pmcid={3287055},
pmid={21968756}}
\bptok{imsref}%
\end{barticle}
%
\endbibitem

\bibitem[\protect\citeauthoryear{Falkner et~al.}{2006}]{Falkner2006}
%
\begin{barticle}[pbm]
\bauthor{\bsnm{Falkner},~\bfnm{Bonita}\binits{B.}},
\bauthor{\bsnm{Gidding},~\bfnm{Samuel~S.}\binits{S.~S.}},
\bauthor{\bsnm{Ramirez-Garnica},~\bfnm{Gabriela}\binits{G.}},
\bauthor{\bsnm{Wiltrout},~\bfnm{Stacey~Armatti}\binits{S.~A.}},
\bauthor{\bsnm{West},~\bfnm{David}\binits{D.}} \AND
\bauthor{\bsnm{Rappaport},~\bfnm{Elizabeth~B.}\binits{E.~B.}}
(\byear{2006}).
\btitle{The relationship of body mass index and blood pressure in
primary care
pediatric patients}.
\bjournal{J. Pediatr.}
\bvolume{148}
\bpages{195--200}.
\bid{doi={10.1016/j.jpeds.2005.10.030}, issn={0022-3476},
pii={S0022-3476(05)01002-4}, pmid={16492428}}
\bptok{imsref}%
\end{barticle}
%
\endbibitem

\bibitem[\protect\citeauthoryear{Fujita et~al.}{2010}]{Fujita2010}
%
\begin{barticle}[pbm]
\bauthor{\bsnm{Fujita},~\bfnm{Yuki}\binits{Y.}},
\bauthor{\bsnm{Kouda},~\bfnm{Katsuyasu}\binits{K.}},
\bauthor{\bsnm{Nakamura},~\bfnm{Harunobu}\binits{H.}},
\bauthor{\bsnm{Nishio},~\bfnm{Nobuhiro}\binits{N.}},
\bauthor{\bsnm{Takeuchi},~\bfnm{Hiroichi}\binits{H.}} \AND
\bauthor{\bsnm{Iki},~\bfnm{Masayuki}\binits{M.}}
(\byear{2010}).
\btitle{Relationship between height and blood pressure in Japanese
schoolchildren}.
\bjournal{Pediatr. Int.}
\bvolume{52}
\bpages{689--693}.
\bid{doi={10.1111/j.1442-200X.2010.03093.x}, issn={1442-200X}, pii={PED3093},
pmid={20136723}}
\bptok{imsref}%
\end{barticle}
%
\endbibitem

\bibitem[\protect\citeauthoryear{Ghosh and Hanson}{2010}]{Ghosh2010}
%
\begin{barticle}[mr]
\bauthor{\bsnm{Ghosh},~\bfnm{Pulak}\binits{P.}} \AND
\bauthor{\bsnm{Hanson},~\bfnm{Timothy}\binits{T.}}
(\byear{2010}).
\btitle{A semiparametric {B}ayesian approach to multivariate longitudinal
data}.
\bjournal{Aust. N. Z. J. Stat.}
\bvolume{52}
\bpages{275--288}.
\bid{doi={10.1111/j.1467-842X.2010.00581.x}, issn={1369-1473}, mr={2744574}}
\bptok{imsref}%
\end{barticle}
%
\endbibitem

\bibitem[\protect\citeauthoryear{Ghosh and Tu}{2009}]{Ghosh2009}
%
\begin{barticle}[mr]
\bauthor{\bsnm{Ghosh},~\bfnm{Pulak}\binits{P.}} \AND
\bauthor{\bsnm{Tu},~\bfnm{Wanzhu}\binits{W.}}
(\byear{2009}).
\btitle{Assessing sexual attitudes and behaviors of young women: A
joint model
with nonlinear time effects, time varying covariates, and dropouts}.
\bjournal{J.~Amer. Statist. Assoc.}
\bvolume{104}
\bpages{474--485}.
\bid{doi={10.1198/jasa.2009.0013}, issn={0162-1459}, mr={2751432}}
\bptok{imsref}%
\end{barticle}
%

\endbibitem
\bibitem[\protect\citeauthoryear{Green and Silverman}{1994}]{Green1994}
%
\begin{bbook}[mr]
\bauthor{\bsnm{Green},~\bfnm{P.~J.}\binits{P.~J.}} \AND
\bauthor{\bsnm{Silverman},~\bfnm{B.~W.}\binits{B.~W.}}
(\byear{1994}).
\btitle{Nonparametric Regression and Generalized Linear Models: A
Roughness Penalty Approach}.
\bseries{Monographs on Statistics and Applied Probability}
\bvolume{58}.
\bpublisher{Chapman \& Hall}, \baddress{London}.
\bid{mr={1270012}}
\bptok{imsref}%
\end{bbook}
%
\endbibitem

\bibitem[\protect\citeauthoryear{Guillas and Lai}{2010}]{Guillas2010}
%
\begin{barticle}[mr]
\bauthor{\bsnm{Guillas},~\bfnm{Serge}\binits{S.}} \AND
\bauthor{\bsnm{Lai},~\bfnm{Ming-Jun}\binits{M.-J.}}
(\byear{2010}).
\btitle{Bivariate splines for spatial functional regression models}.
\bjournal{J.~Nonparametr. Stat.}
\bvolume{22}
\bpages{477--497}.
\bid{doi={10.1080/10485250903323180}, issn={1048-5252}, mr={2662608}}
\bptok{imsref}%
\end{barticle}
%
\endbibitem

\bibitem[\protect\citeauthoryear{Guo and Carlin}{2004}]{Guo2004}
%
\begin{barticle}[mr]
\bauthor{\bsnm{Guo},~\bfnm{Xu}\binits{X.}} \AND
\bauthor{\bsnm{Carlin},~\bfnm{Bradley~P.}\binits{B.~P.}}
(\byear{2004}).
\btitle{Separate and joint modeling of longitudinal and event time data using
standard computer packages}.
\bjournal{Amer. Statist.}
\bvolume{58}
\bpages{16--24}.
\bid{doi={10.1198/0003130042854}, issn={0003-1305}, mr={2055507}}
\bptok{imsref}%
\end{barticle}
%
\endbibitem

\bibitem[\protect\citeauthoryear{Hall}{2003}]{Hall2003}
%
\begin{barticle}[pbm]
\bauthor{\bsnm{Hall},~\bfnm{John~E.}\binits{J.~E.}}
(\byear{2003}).
\btitle{The kidney, hypertension, and obesity}.
\bjournal{Hypertension}
\bvolume{41}
\bpages{625--633}.
\bid{doi={10.1161/01.HYP.0000052314.95497.78}, issn={1524-4563},
pii={01.HYP.0000052314.95497.78}, pmid={12623970}}
\bptok{imsref}%
\end{barticle}
%
\endbibitem

\bibitem[\protect\citeauthoryear{Hall et~al.}{2010}]{Hall2010}
%
\begin{barticle}[pbm]
\bauthor{\bsnm{Hall},~\bfnm{John~E.}\binits{J.~E.}}, \bauthor{\bparticle{da}
\bsnm{Silva},~\bfnm{Alexandre~A.}\binits{A.~A.}}, \bauthor{\bparticle{do}
\bsnm{Carmo},~\bfnm{Jussara~M.}\binits{J.~M.}},
\bauthor{\bsnm{Dubinion},~\bfnm{John}\binits{J.}},
\bauthor{\bsnm{Hamza},~\bfnm{Shereen}\binits{S.}},
\bauthor{\bsnm{Munusamy},~\bfnm{Shankar}\binits{S.}},
\bauthor{\bsnm{Smith},~\bfnm{Grant}\binits{G.}} \AND
\bauthor{\bsnm{Stec},~\bfnm{David~E.}\binits{D.~E.}}
(\byear{2010}).
\btitle{Obesity-induced hypertension: Role of sympathetic nervous system,
leptin, and melanocortins}.
\bjournal{J. Biol. Chem.}
\bvolume{285}
\bpages{17271--17276}.
\bid{doi={10.1074/jbc.R110.113175}, issn={1083-351X}, pii={R110.113175},
pmcid={2878489}, pmid={20348094}}
\bptok{imsref}%
\end{barticle}
%
\endbibitem

\bibitem[\protect\citeauthoryear{He, Fung and Zhu}{2005}]{He2005}
%
\begin{barticle}[mr]
\bauthor{\bsnm{He},~\bfnm{Xuming}\binits{X.}},
\bauthor{\bsnm{Fung},~\bfnm{Wing~K.}\binits{W.~K.}} \AND
\bauthor{\bsnm{Zhu},~\bfnm{Zhongyi}\binits{Z.}}
(\byear{2005}).
\btitle{Robust estimation in generalized partial linear models for clustered
data}.
\bjournal{J. Amer. Statist. Assoc.}
\bvolume{100}
\bpages{1176--1184}.
\bid{doi={10.1198/016214505000000277}, issn={0162-1459}, mr={2236433}}
\bptok{imsref}%
\end{barticle}
%
\endbibitem

\bibitem[\protect\citeauthoryear{Huang et~al.}{1998}]{Huang1998}
%
\begin{barticle}[author]
\bauthor{\bsnm{Huang},~\bfnm{Z.}\binits{Z.}},
\bauthor{\bsnm{Willett},~\bfnm{W.~C.}\binits{W.~C.}},
\bauthor{\bsnm{Manson},~\bfnm{J.~E.}\binits{J.~E.}},
\bauthor{\bsnm{Rosner},~\bfnm{B.}\binits{B.}},
\bauthor{\bsnm{Stampfer},~\bfnm{M.~J.}\binits{M.~J.}},
\bauthor{\bsnm{Speizer},~\bfnm{F.~E.}\binits{F.~E.}} \AND
\bauthor{\bsnm{Colditz},~\bfnm{G.~A.}\binits{G.~A.}}
(\byear{1998}).
\btitle{Body weight, weight change, and risk for hypertension in women}.
\bjournal{Annals of Internal Medicine}
\bvolume{128}
\bpages{81--88}.
\bptok{imsref}%
\end{barticle}
%
\endbibitem

\bibitem[\protect\citeauthoryear{Humphreys}{2011}]{Humphreys2011}
%
\begin{barticle}[pbm]
\bauthor{\bsnm{Humphreys},~\bfnm{Michael~H.}\binits{M.~H.}}
(\byear{2011}).
\btitle{The brain splits obesity and hypertension}.
\bjournal{Nat. Med.}
\bvolume{17}
\bpages{782--783}.
\bid{doi={10.1038/nm0711-782}, issn={1546-170X}, pii={nm0711-782},
pmid={21738154}}
\bptok{imsref}%
\end{barticle}
%
\endbibitem

\bibitem[\protect\citeauthoryear{Jones}{1993}]{Jones1993}
%
\begin{bbook}[mr]
\bauthor{\bsnm{Jones},~\bfnm{R.~H.}\binits{R.~H.}}
(\byear{1993}).
\btitle{Longitudinal Data with Serial Correlation: A State-Space Approach}.
\bseries{Monographs on Statistics and Applied Probability}
\bvolume{47}.
\bpublisher{Chapman \& Hall}, \baddress{London}.
\bid{mr={1293123}}
\bptok{imsref}%
\end{bbook}
%
\endbibitem

\bibitem[\protect\citeauthoryear{Kohn, Ansley and Tharm}{1991}]{Kohn1991}
%
\begin{barticle}[mr]
\bauthor{\bsnm{Kohn},~\bfnm{Robert}\binits{R.}},
\bauthor{\bsnm{Ansley},~\bfnm{Craig~F.}\binits{C.~F.}} \AND
\bauthor{\bsnm{Tharm},~\bfnm{David}\binits{D.}}
(\byear{1991}).
\btitle{The performance of cross-validation and maximum likelihood estimators
of spline smoothing parameters}.
\bjournal{J. Amer. Statist. Assoc.}
\bvolume{86}
\bpages{1042--1050}.
\bid{issn={0162-1459}, mr={1146351}}
\bptok{imsref}%
\end{barticle}
%
\endbibitem

\bibitem[\protect\citeauthoryear{Lauer and Clarke}{1989}]{Lauer1989}
%
\begin{barticle}[pbm]
\bauthor{\bsnm{Lauer},~\bfnm{R.~M.}\binits{R.~M.}} \AND
\bauthor{\bsnm{Clarke},~\bfnm{W.~R.}\binits{W.~R.}}
(\byear{1989}).
\btitle{Childhood risk factors for high adult blood pressure: The
muscatine study}.
\bjournal{Pediatrics}
\bvolume{84}
\bpages{633--641}.
\bid{issn={0031-4005}, pmid={2780125}}
\bptok{imsref}%
\end{barticle}
%
\endbibitem

\bibitem[\protect\citeauthoryear{Lever and Harrap}{1992}]{Lever1992}
%
\begin{barticle}[pbm]
\bauthor{\bsnm{Lever},~\bfnm{A.~F.}\binits{A.~F.}} \AND
\bauthor{\bsnm{Harrap},~\bfnm{S.~B.}\binits{S.~B.}}
(\byear{1992}).
\btitle{Essential hypertension: A disorder of growth with origins in
childhood?}
\bjournal{J. Hypertens.}
\bvolume{10}
\bpages{101--120}.
\bid{issn={0263-6352}, pmid={1313473}}
\bptok{imsref}%
\end{barticle}
%
\endbibitem

\bibitem[\protect\citeauthoryear{Levin et~al.}{2010}]{Levin2010}
%
\begin{barticle}[pbm]
\bauthor{\bsnm{Levin},~\bfnm{Avi}\binits{A.}},
\bauthor{\bsnm{Morad},~\bfnm{Yair}\binits{Y.}},
\bauthor{\bsnm{Grotto},~\bfnm{Itamar}\binits{I.}},
\bauthor{\bsnm{Ravid},~\bfnm{Mordechai}\binits{M.}} \AND
\bauthor{\bsnm{Bar-Dayan},~\bfnm{Yosefa}\binits{Y.}}
(\byear{2010}).
\btitle{Weight disorders and associated morbidity among young adults in Israel
1990--2003}.
\bjournal{Pediatr. Int.}
\bvolume{52}
\bpages{347--352}.
\bid{doi={10.1111/j.1442-200X.2009.02972.x}, issn={1442-200X}, pii={PED2972},
pmid={19807878}}
\bptok{imsref}%
\end{barticle}
%
\endbibitem

\bibitem[\protect\citeauthoryear{Lin and Carroll}{2006}]{Lin2006}
%
\begin{barticle}[mr]
\bauthor{\bsnm{Lin},~\bfnm{Xihong}\binits{X.}} \AND
\bauthor{\bsnm{Carroll},~\bfnm{Raymond~J.}\binits{R.~J.}}
(\byear{2006}).
\btitle{Semiparametric estimation in general repeated measures problems}.
\bjournal{J. R. Stat. Soc. Ser. B Stat. Methodol.}
\bvolume{68}
\bpages{69--88}.
\bid{doi={10.1111/j.1467-9868.2005.00533.x}, issn={1369-7412}, mr={2212575}}
\bptok{imsref}%
\end{barticle}
%
\endbibitem

\bibitem[\protect\citeauthoryear{Lin and Zhang}{1999}]{Lin1999}
%
\begin{barticle}[mr]
\bauthor{\bsnm{Lin},~\bfnm{Xihong}\binits{X.}} \AND
\bauthor{\bsnm{Zhang},~\bfnm{Daowen}\binits{D.}}
(\byear{1999}).
\btitle{Inference in generalized additive mixed models by using smoothing
splines}.
\bjournal{J. R. Stat. Soc. Ser. B Stat. Methodol.}
\bvolume{61}
\bpages{381--400}.
\bid{doi={10.1111/1467-9868.00183}, issn={1369-7412}, mr={1680318}}
\bptok{imsref}%
\end{barticle}
%
\endbibitem

\bibitem[\protect\citeauthoryear{Liu}{1988}]{Liu1988}
%
\begin{barticle}[mr]
\bauthor{\bsnm{Liu},~\bfnm{Regina~Y.}\binits{R.~Y.}}
(\byear{1988}).
\btitle{Bootstrap procedures under some non-i.i.d. models}.
\bjournal{Ann. Statist.}
\bvolume{16}
\bpages{1696--1708}.
\bid{doi={10.1214/aos/1176351062}, issn={0090-5364}, mr={0964947}}
\bptok{imsref}%
\end{barticle}
%
\endbibitem

\bibitem[\protect\citeauthoryear{Liu and Tu}{2012}]{LiuTu2012S}
%
\begin{bmisc}[author]
\bauthor{\bsnm{Liu},~\bfnm{H.}\binits{H.}} \AND
\bauthor{\bsnm{Tu},~\bfnm{W.}\binits{W.}}
(\byear{2012}).
\bhowpublished{Supplement to ``A semiparametric regression model for
paired longitudinal outcomes
with application in childhood blood pressure development.'' DOI:\doiurl
{10.1214/12-AOAS567SUPP}.}
\bptok{imsref}%
\end{bmisc}
%
\endbibitem

\bibitem[\protect\citeauthoryear{Mammen}{1993}]{Mammen1993}
%
\begin{barticle}[mr]
\bauthor{\bsnm{Mammen},~\bfnm{Enno}\binits{E.}}
(\byear{1993}).
\btitle{Bootstrap and wild bootstrap for high-dimensional linear models}.
\bjournal{Ann. Statist.}
\bvolume{21}
\bpages{255--285}.
\bid{doi={10.1214/aos/1176349025}, issn={0090-5364}, mr={1212176}}
\bptok{imsref}%
\end{barticle}
%
\endbibitem

\bibitem[\protect\citeauthoryear{Masuo et~al.}{2000}]{Masuo2000}
%
\begin{barticle}[pbm]
\bauthor{\bsnm{Masuo},~\bfnm{K.}\binits{K.}},
\bauthor{\bsnm{Mikami},~\bfnm{H.}\binits{H.}},
\bauthor{\bsnm{Ogihara},~\bfnm{T.}\binits{T.}} \AND
\bauthor{\bsnm{Tuck},~\bfnm{M.~L.}\binits{M.~L.}}
(\byear{2000}).
\btitle{Weight gain-induced blood pressure elevation}.
\bjournal{Hypertension}
\bvolume{35}
\bpages{1135--1140}.
\bid{issn={1524-4563}, pmid={10818077}}
\bptok{imsref}%
\end{barticle}
%
\endbibitem

\bibitem[\protect\citeauthoryear{Morris}{2002}]{Morris2002}
%
\begin{barticle}[mr]
\bauthor{\bsnm{Morris},~\bfnm{Jeffrey~S.}\binits{J.~S.}}
(\byear{2002}).
\btitle{The {BLUP}s are not ``best'' when it comes to bootstrapping}.
\bjournal{Statist. Probab. Lett.}
\bvolume{56}
\bpages{425--430}.
\bid{doi={10.1016/S0167-7152(02)00041-X}, issn={0167-7152}, mr={1898721}}
\bptok{imsref}%
\end{barticle}
%
\endbibitem

\bibitem[\protect\citeauthoryear{Morris et~al.}{2001}]{Morris2001}
%
\begin{barticle}[mr]
\bauthor{\bsnm{Morris},~\bfnm{Jeffrey~S.}\binits{J.~S.}},
\bauthor{\bsnm{Wang},~\bfnm{Naisyin}\binits{N.}},
\bauthor{\bsnm{Lupton},~\bfnm{Joanne~R.}\binits{J.~R.}},
\bauthor{\bsnm{Chapkin},~\bfnm{Robert~S.}\binits{R.~S.}},
\bauthor{\bsnm{Turner},~\bfnm{Nancy~D.}\binits{N.~D.}},
\bauthor{\bsnm{Hong},~\bfnm{Mee~Young}\binits{M.~Y.}} \AND
\bauthor{\bsnm{Carroll},~\bfnm{Raymond~J.}\binits{R.~J.}}
(\byear{2001}).
\btitle{Parametric and nonparametric methods for understanding the relationship
between carcinogen-induced {DNA} adduct levels in distal and proximal regions
of the colon}.
\bjournal{J. Amer. Statist. Assoc.}
\bvolume{96}
\bpages{816--826}.
\bid{doi={10.1198/016214501753208528}, issn={0162-1459}, mr={1946358}}
\bptok{imsref}%
\end{barticle}
%
\endbibitem

\bibitem[\protect\citeauthoryear{Pickering et~al.}{2005}]{Pickering2005}
%
\begin{barticle}[pbm]
\bauthor{\bsnm{Pickering},~\bfnm{Thomas~G.}\binits{T.~G.}},
\bauthor{\bsnm{Hall},~\bfnm{John~E.}\binits{J.~E.}},
\bauthor{\bsnm{Appel},~\bfnm{Lawrence~J.}\binits{L.~J.}},
\bauthor{\bsnm{Falkner},~\bfnm{Bonita~E.}\binits{B.~E.}},
\bauthor{\bsnm{Graves},~\bfnm{John}\binits{J.}},
\bauthor{\bsnm{Hill},~\bfnm{Martha~N.}\binits{M.~N.}},
\bauthor{\bsnm{Jones},~\bfnm{Daniel~W.}\binits{D.~W.}},
\bauthor{\bsnm{Kurtz},~\bfnm{Theodore}\binits{T.}},
\bauthor{\bsnm{Sheps},~\bfnm{Sheldon~G.}\binits{S.~G.}}
\AND
\bauthor{\bsnm{Roccella},~\bfnm{Edward~J.}\binits{E.~J.}}
(\byear{2005}).
\btitle{Recommendations for blood pressure measurement in humans and
experimental animals: Part 1: Blood pressure measurement in humans: A
statement for professionals from the Subcommittee of Professional and Public
Education of the American Heart Association Council on High Blood Pressure
Research}.
\bjournal{Hypertension}
\bvolume{45}
\bpages{142--161}.
\bid{doi={10.1161/01.HYP.0000150859.47929.8e}, issn={1524-4563},
pii={01.HYP.0000150859.47929.8e}, pmid={15611362}}
\bptok{imsref}%
\end{barticle}
%
\endbibitem

\bibitem[\protect\citeauthoryear{Pinheiro and Bates}{2000}]{Pinheiro2000}
%
\begin{bbook}[author]
\bauthor{\bsnm{Pinheiro},~\bfnm{J.~C.}\binits{J.~C.}} \AND
\bauthor{\bsnm{Bates},~\bfnm{D.~M.}\binits{D.~M.}}
(\byear{2000}).
\btitle{Mixed-Effects Models in S and S-PLUS}.
\bpublisher{Springer}, \baddress{New York}.
\bptok{imsref}%
\end{bbook}
%
\endbibitem

\bibitem[\protect\citeauthoryear{Purkayastha, Zhang and
Cai}{2011}]{Purkayastha2011}
%
\begin{barticle}[pbm]
\bauthor{\bsnm{Purkayastha},~\bfnm{Sudarshana}\binits{S.}},
\bauthor{\bsnm{Zhang},~\bfnm{Guo}\binits{G.}} \AND
\bauthor{\bsnm{Cai},~\bfnm{Dongsheng}\binits{D.}}
(\byear{2011}).
\btitle{Uncoupling the mechanisms of obesity and hypertension by targeting
hypothalamic IKK-beta and NF-kappa B}.
\bjournal{Nat. Med.}
\bvolume{17}
\bpages{883--887}.
\bid{doi={10.1038/nm.2372}, issn={1546-170X}, mid={NIHMS286510}, pii={nm.2372},
pmcid={3134198}, pmid={21642978}}
\bptok{imsref}%
\end{barticle}
%
\endbibitem

\bibitem[\protect\citeauthoryear{Roca-Pardi{\~n}as et~al.}{2008}]{Roca2008}
%
\begin{barticle}[mr]
\bauthor{\bsnm{Roca-Pardi{\~n}as},~\bfnm{Javier}\binits{J.}},
\bauthor{\bsnm{Cadarso-Su{\'a}rez},~\bfnm{Carmen}\binits{C.}},
\bauthor{\bsnm{Tahoces},~\bfnm{Pablo~G.}\binits{P.~G.}} \AND
\bauthor{\bsnm{Lado},~\bfnm{Mar{\'{\i}}a~J.}\binits{M.~J.}}
(\byear{2008}).
\btitle{Assessing continuous bivariate effects among different groups through
nonparametric regression models: An application to breast cancer detection}.
\bjournal{Comput. Statist. Data Anal.}
\bvolume{52}
\bpages{1958--1970}.
\bid{doi={10.1016/j.csda.2007.06.024}, issn={0167-9473}, mr={2418483}}
\bptok{imsref}%
\end{barticle}
%
\endbibitem

\bibitem[\protect\citeauthoryear{Ruppert, Wand and Carroll}{2003}]{Ruppert2003}
%
\begin{bbook}[mr]
\bauthor{\bsnm{Ruppert},~\bfnm{David}\binits{D.}},
\bauthor{\bsnm{Wand},~\bfnm{M.~P.}\binits{M.~P.}} \AND
\bauthor{\bsnm{Carroll},~\bfnm{R.~J.}\binits{R.~J.}}
(\byear{2003}).
\btitle{Semiparametric Regression}.
\bseries{Cambridge Series in Statistical and Probabilistic Mathematics}
\bvolume{12}.
\bpublisher{Cambridge Univ. Press}, \baddress{Cambridge}.
\bid{doi={10.1017/CBO9780511755453}, mr={1998720}}
\bptok{imsref}%
\end{bbook}
%
\endbibitem

\bibitem[\protect\citeauthoryear{Sain et~al.}{2006}]{Sain2006}
%
\begin{barticle}[author]
\bauthor{\bsnm{Sain},~\bfnm{S.~R.}\binits{S.~R.}},
\bauthor{\bsnm{Jagtap},~\bfnm{S.}\binits{S.}},
\bauthor{\bsnm{Mearns},~\bfnm{L.}\binits{L.}} \AND
\bauthor{\bsnm{Nychka},~\bfnm{D.}\binits{D.}}
(\byear{2006}).
\btitle{A multivariate spatial model for soil water profiles}.
\bjournal{J. Agric. Biol. Environ. Stat.}
\bvolume{11}
\bpages{462--480}.
\bptok{imsref}%
\end{barticle}
%
\endbibitem

\bibitem[\protect\citeauthoryear{Self and Liang}{1987}]{SelfLiang1987}
%
\begin{barticle}[mr]
\bauthor{\bsnm{Self},~\bfnm{Steven~G.}\binits{S.~G.}} \AND
\bauthor{\bsnm{Liang},~\bfnm{Kung-Yee}\binits{K.-Y.}}
(\byear{1987}).
\btitle{Asymptotic properties of maximum likelihood estimators and likelihood
ratio tests under nonstandard conditions}.
\bjournal{J. Amer. Statist. Assoc.}
\bvolume{82}
\bpages{605--610}.
\bid{issn={0162-1459}, mr={0898365}}
\bptok{imsref}%
\end{barticle}
%
\endbibitem

\bibitem[\protect\citeauthoryear{Shankar et~al.}{2005}]{Shankar2005}
%
\begin{barticle}[author]
\bauthor{\bsnm{Shankar},~\bfnm{R.~R.}\binits{R.~R.}},
\bauthor{\bsnm{Eckert},~\bfnm{G.~J.}\binits{G.~J.}},
\bauthor{\bsnm{Saha},~\bfnm{C.}\binits{C.}},
\bauthor{\bsnm{Tu},~\bfnm{W.}\binits{W.}} \AND
\bauthor{\bsnm{Pratt},~\bfnm{J.~H.}\binits{J.~H.}}
(\byear{2005}).
\btitle{The change in blood pressure during pubertal growth}.
\bjournal{Journal of Clinical Endocrinology \& Metabolism}
\bvolume{90}
\bpages{163--167}.
\bptok{imsref}%
\end{barticle}
%
\endbibitem

\bibitem[\protect\citeauthoryear{Steinberger et~al.}{2009}]{Steinberger2009}
%
\begin{barticle}[author]
\bauthor{\bsnm{Steinberger},~\bfnm{J.}\binits{J.}},
\bauthor{\bsnm{Daniels},~\bfnm{S.~R.}\binits{S.~R.}},
\bauthor{\bsnm{Eckel},~\bfnm{R.~H.}\binits{R.~H.}},
\bauthor{\bsnm{Hayman},~\bfnm{L.}\binits{L.}},
\bauthor{\bsnm{Lustig},~\bfnm{R.~H.}\binits{R.~H.}},
\bauthor{\bsnm{\mbox{McCrindle}},~\bfnm{B.}\binits{B.}} \AND
\bauthor{\bsnm{Mietus-Snyder},~\bfnm{M.~L.}\binits{M.~L.}}
(\byear{2009}).
\btitle{Progress and challenges in metabolic syndrome in children and
adolescents: A scientific statement from the American Heart Association
atherosclerosis, hypertension, and obesity in the young Committee of the
Council on cardiovascular disease in the young; Council on cardiovascular
nursing; and Council on nutrition, physical activity, and metabolism}.
\bjournal{Circulation}
\bvolume{119}
\bpages{628--647}.
\bptok{imsref}%
\end{barticle}
%
\endbibitem

\bibitem[\protect\citeauthoryear{Stram and Lee}{1994}]{Stram1994}
%
\begin{barticle}[author]
\bauthor{\bsnm{Stram},~\bfnm{D.~O.}\binits{D.~O.}} \AND
\bauthor{\bsnm{Lee},~\bfnm{J.~W.}\binits{J.~W.}}
(\byear{1994}).
\btitle{Variance-components testing in the longitudinal mixed effects model}.
\bjournal{Biometrics}
\bvolume{50}
\bpages{1171--1177}.
\bptok{imsref}%
\end{barticle}
%
\endbibitem

\bibitem[\protect\citeauthoryear{Stray-Pedersen et~al.}{2009}]{Stray2009}
%
\begin{barticle}[pbm]
\bauthor{\bsnm{Stray-Pedersen},~\bfnm{Marit}\binits{M.}},
\bauthor{\bsnm{Helsing},~\bfnm{Ragnhild~M.}\binits{R.~M.}},
\bauthor{\bsnm{Gibbons},~\bfnm{Luz}\binits{L.}},
\bauthor{\bsnm{Cormick},~\bfnm{Gabriela}\binits{G.}},
\bauthor{\bsnm{Holmen},~\bfnm{Turid~L.}\binits{T.~L.}},
\bauthor{\bsnm{Vik},~\bfnm{Torstein}\binits{T.}} \AND
\bauthor{\bsnm{Beliz{\'{a}}n},~\bfnm{Jos{\'{e}}~M.}\binits{J.~M.}}
(\byear{2009}).
\btitle{Weight status and hypertension among adolescent girls in
Argentina and
Norway: Data from the ENNyS and HUNT studies}.
\bjournal{BMC Public Health}
\bvolume{9}
\bpages{398}.
\bid{doi={10.1186/1471-2458-9-398}, issn={1471-2458}, pii={1471-2458-9-398},
pmcid={2775744}, pmid={19878550}}
\bptok{imsref}%
\end{barticle}
%
\endbibitem

\bibitem[\protect\citeauthoryear{Tu et~al.}{2009}]{Tu2009}
%
\begin{barticle}[author]
\bauthor{\bsnm{Tu},~\bfnm{W.}\binits{W.}},
\bauthor{\bsnm{Eckert},~\bfnm{G.~J.}\binits{G.~J.}},
\bauthor{\bsnm{Saha},~\bfnm{C.}\binits{C.}} \AND
\bauthor{\bsnm{Pratt},~\bfnm{J.~H.}\binits{J.~H.}}
(\byear{2009}).
\btitle{Synchronization of adolescent blood pressure and pubertal somatic
growth}.
\bjournal{Journal of Clinical Endocrinology \& Metabolism}
\bvolume{94}
\bpages{5019--5022}.
\bptok{imsref}%
\end{barticle}
%
\endbibitem

\bibitem[\protect\citeauthoryear{Tu et~al.}{2011}]{Tu2011}
%
\begin{barticle}[author]
\bauthor{\bsnm{Tu},~\bfnm{W}\binits{W.}},
\bauthor{\bsnm{Eckert},~\bfnm{G.~J.}\binits{G.~J.}},
\bauthor{\bsnm{DiMeglio},~\bfnm{L.~A.}\binits{L.~A.}},
\bauthor{\bsnm{Yu},~\bfnm{Z.}\binits{Z.}},
\bauthor{\bsnm{Jung},~\bfnm{J.}\binits{J.}} \AND
\bauthor{\bsnm{Pratt},~\bfnm{J.~H.}\binits{J.~H.}}
(\byear{2011}).
\btitle{Intensified effect of adiposity on blood pressure in overweight and
obese children}.
\bjournal{Hypertension}
\bvolume{58}
\bpages{818--824}.
\bptok{imsref}%
\end{barticle}
%
\endbibitem

\bibitem[\protect\citeauthoryear{Wahba}{1985}]{Wahba1985}
%
\begin{barticle}[mr]
\bauthor{\bsnm{Wahba},~\bfnm{Grace}\binits{G.}}
(\byear{1985}).
\btitle{A comparison of {GCV} and {GML} for choosing the smoothing
parameter in
the generalized spline smoothing problem}.
\bjournal{Ann. Statist.}
\bvolume{13}
\bpages{1378--1402}.
\bid{doi={10.1214/aos/1176349743}, issn={0090-5364}, mr={0811498}}
\bptok{imsref}%
\end{barticle}
%
\endbibitem

\bibitem[\protect\citeauthoryear{Wood}{2003}]{Wood2003}
%
\begin{barticle}[mr]
\bauthor{\bsnm{Wood},~\bfnm{Simon~N.}\binits{S.~N.}}
(\byear{2003}).
\btitle{Thin plate regression splines}.
\bjournal{J. R. Stat. Soc. Ser. B Stat. Methodol.}
\bvolume{65}
\bpages{95--114}.
\bid{doi={10.1111/1467-9868.00374}, issn={1369-7412}, mr={1959095}}
\bptok{imsref}%
\end{barticle}
%
\endbibitem

\bibitem[\protect\citeauthoryear{Wood}{2006}]{Wood2006}
%
\begin{bbook}[mr]
\bauthor{\bsnm{Wood},~\bfnm{Simon~N.}\binits{S.~N.}}
(\byear{2006}).
\btitle{Generalized Additive Models: An Introduction with $R$}.
\bpublisher{Chapman \& Hall/CRC}, \baddress{Boca Raton, FL}.
\bid{mr={2206355}}
\bptok{imsref}%
\end{bbook}
%
\endbibitem

\bibitem[\protect\citeauthoryear{Wood}{2010}]{mgcv}
%
\begin{bmisc}[author]
\bauthor{\bsnm{Wood},~\bfnm{S.~N.}\binits{S.~N.}}
(\byear{2010}).
\bhowpublished{mgcv: GAMs with GCV/AIC/REML smoothness estimation and
GAMMs by PQL.
R package version 1.7-2.}
\bptok{imsref}%
\end{bmisc}
%
\endbibitem

\bibitem[\protect\citeauthoryear{Zhang and Lin}{2003}]{Zhang2003}
%
\begin{barticle}[author]
\bauthor{\bsnm{Zhang},~\bfnm{D.}\binits{D.}} \AND
\bauthor{\bsnm{Lin},~\bfnm{X.}\binits{X.}}
(\byear{2003}).
\btitle{Hypothesis testing in semiparametric additive mixed models}.
\bjournal{Biostatistics}
\bvolume{4}
\bpages{57--74}.
\bptok{imsref}%
\end{barticle}
%
\endbibitem

\end{thebibliography}
\end{document}